\documentclass[acmtog, nonacm]{acmart}

\usepackage{float}
\usepackage{tabularx}
\usepackage{soul}
\usepackage{comment}
\usepackage{booktabs}
\AtBeginDocument{%
  }

\setcopyright{acmlicensed}
\copyrightyear{2025}
\acmYear{2025}
\acmDOI{XXXXXXX.XXXXXXX}

\acmJournal{TOG}
\acmVolume{}
\acmNumber{}
\acmArticle{}
\acmMonth{}

\begin{document}


\title{Security Vulnerabilities in Software Supply Chain for Autonomous Vehicles}

\author{Md Wasiul Haque}
\affiliation{%
  \institution{Department of Civil, Construction and Environmental Engineering, University of Alabama}
  \city{Tuscaloosa}
  \state{AL}
  \country{USA}}
\email{mdhaque@ua.edu} 

\author{Md Erfan}
\affiliation{%
  \institution{Department of Computer Science, University of Alabama}
  \city{Tuscaloosa}
  \state{AL}
  \country{USA}}
\email{merfan@ua.edu}

\author{Sagar Dasgupta}
\affiliation{%
  \institution{Department of Civil, Construction and Environmental Engineering, University of Alabama}
  \city{Tuscaloosa}
  \state{AL}
  \country{USA}}
\email{sagar@ua.edu}

\author{Md Rayhanur Rahman}
\affiliation{%
  \institution{Department of Computer Science, University of Alabama}
  \city{Tuscaloosa}
  \state{AL}
  \country{USA}}
\email{rayhanur@ua.edu}

\author{Mizanur Rahman}
\affiliation{%
  \institution{Department of Civil, Construction and Environmental Engineering, University of Alabama}
  \city{Tuscaloosa}
  \state{AL}
  \country{USA}}
\email{mrahman@ua.edu}

\renewcommand{\shortauthors}{Haque et al.}

\begin{abstract}
The interest in autonomous vehicles (AVs) for critical missions, including transportation, rescue, surveillance, reconnaissance, and mapping, is growing rapidly due to their significant safety and mobility benefits. AVs consist of complex software systems that leverage artificial intelligence (AI), sensor fusion algorithms, and real-time data processing. Additionally, AVs are becoming increasingly reliant on open-source software supply chains, such as open-source packages, third-party software components, AI models, and third-party datasets. Software security best practices in the automotive sector are often an afterthought for developers. Thus, significant cybersecurity risks exist in the software supply chain of AVs, particularly when secure software development practices are not rigorously implemented. For example, Upstream’s 2024 Automotive Cybersecurity Report states that 49.5\% of cyberattacks in the automotive sector are related to exploiting security vulnerabilities in software systems. In this chapter, we analyze security vulnerabilities in open-source software components in AVs. We utilize static analyzers on popular open-source AV software, such as Autoware, Apollo, and openpilot.
Specifically, this chapter covers: (1) prevalent software security vulnerabilities of AVs; and (2) a comparison of static analyzer outputs for different open-source AV repositories. The goal is to inform researchers, practitioners, and policymakers about the existing security flaws in the commonplace open-source software ecosystem in the AV domain. The findings would emphasize the necessity of security best practices earlier in the software development lifecycle to reduce cybersecurity risks, thereby ensuring system reliability, safeguarding user data, and maintaining public trust in an increasingly automated world.
\end{abstract}




\maketitle

\section{Introduction}
The rise of autonomous vehicles (AVs) represents one of the most profound technological shifts in modern transportation. Autonomous vehicles integrate perception sensors, precise localization and mapping, trajectory prediction, planning and control systems, real-time decision-making frameworks, and built-in safety redundancies to reduce human error, enhance mobility, and reshape urban ecosystems. \cite{shladover2018connected, liu2021decision}. As vehicles increasingly evolve into software-defined systems, much of their safety, performance, and functionality is dictated not only by mechanical design but also by the reliability and integrity of complex software stacks. This transformation, while enabling unprecedented flexibility and innovation, simultaneously exposes AVs to new classes of risks: software defects, insecure integration of third-party components, and vulnerabilities across the software supply chain \cite{ENISA2021, NIST2022}. 

\begin{figure}[H]
    \centering
    \includegraphics[width=\linewidth]{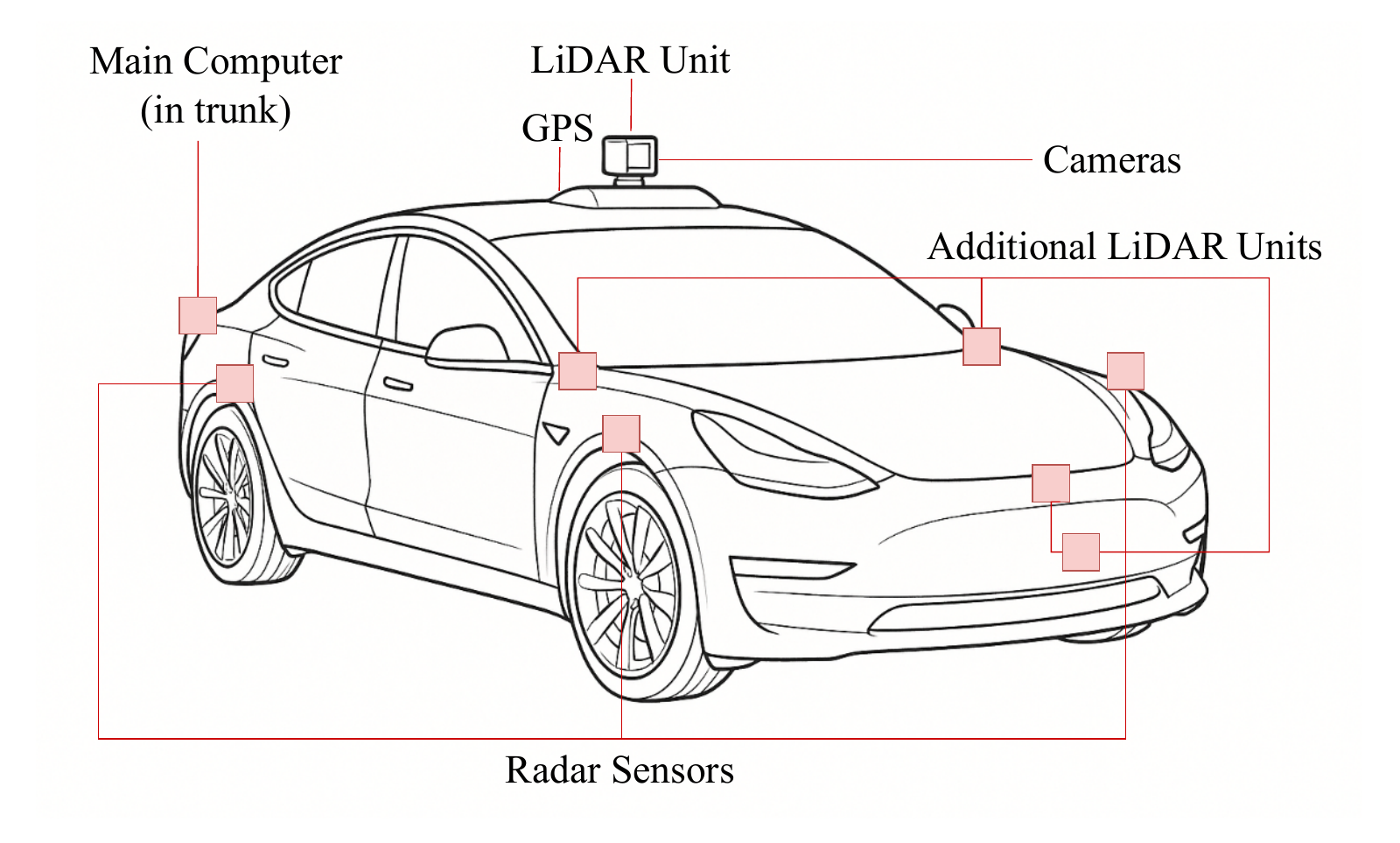}
    \caption{Typical sensor and processing layout of an autonomous vehicle}
    \label{fig:av_graphics}
\end{figure}

Unlike conventional vehicles, AVs are tightly coupled cyber-physical systems, where flaws in the software supply chain can manifest directly as a safety hazard on the road. For instance, compromised dependencies, misconfigured continuous integration (CI) pipelines, or malicious open-source packages may cascade from perception modules into planning and control, resulting in unsafe maneuvers and potential accidents \cite{ladisa2023sok, guo2023empirical}. Reports already indicate that nearly half of all cyberattacks in the automotive sector exploit weaknesses in software systems, underscoring the urgency of securing AV supply chains \cite{durlik2024cybersecurity}. At the same time, the AV domain's reliance on both open-source ecosystems (e.g., Autoware, Apollo, openpilot) and proprietary platforms highlights the trade-offs between transparency, innovation, and the controlled assurance offered by closed industrial systems \cite{kato2018autoware, yurtsever2020survey, aliane2025survey}.

This chapter addresses these challenges by analyzing software supply chain vulnerabilities in AVs through a multi-layered approach. We first examine the architecture of AVs as software-defined vehicles and identify how dependencies across hardware and software amplify supply chain risks. We then review existing vulnerability analysis methods, threat modeling tools, and standard risk assessment approaches \cite{ISO21434}, situating them in the context of safety-critical automotive systems. Building on this foundation, we apply static code analyzers to mainstream open-source AV repositories to uncover Common Weakness Enumerations (CWEs) within their code bases, and complement this with Software Bill of Materials (SBOM) tools to detect vulnerable third-party dependencies. Finally, we explore how these findings can be translated into meaningful risk scoring that reflects both technical severity and system-level safety impact \cite{cvss4}.

By combining a structured review with empirical vulnerability analysis of widely used open-source AV software stacks, this work seeks to inform researchers, practitioners, and policymakers about the real and emerging risks in AV software supply chains. Ultimately, the goal is to emphasize the importance of embedding secure development practices early in the lifecycle of AV software, thereby enhancing resilience, maintaining public trust, and ensuring that the promise of autonomous mobility can be realized safely.

\section{Autonomous Vehicles: Taxonomy, System Architecture, and Software Foundation}
AVs represent a paradigm shift in transportation, offering the potential to fundamentally change how people and goods move within modern societies. Their popularity stems from the potential to mitigate human error, while improving traffic efficiency, reducing congestion, and enhancing mobility for individuals facing difficulty driving, such as the elderly or disabled \cite{litman2017autonomous}. Compared to conventional vehicles, AVs promise lower operational costs, increased fuel efficiency through optimized driving patterns, and environmental benefits from reduced emissions \cite{fagnant2015preparing}. In addition, the socio-economic potential of AVs is vast, ranging from rephrasing urban planning to enabling innovative mobility services such as autonomous taxis and freight delivery systems \cite{anderson2014autonomous}.

\begin{figure}[htbp]
    \centering
    \includegraphics[width=\linewidth]{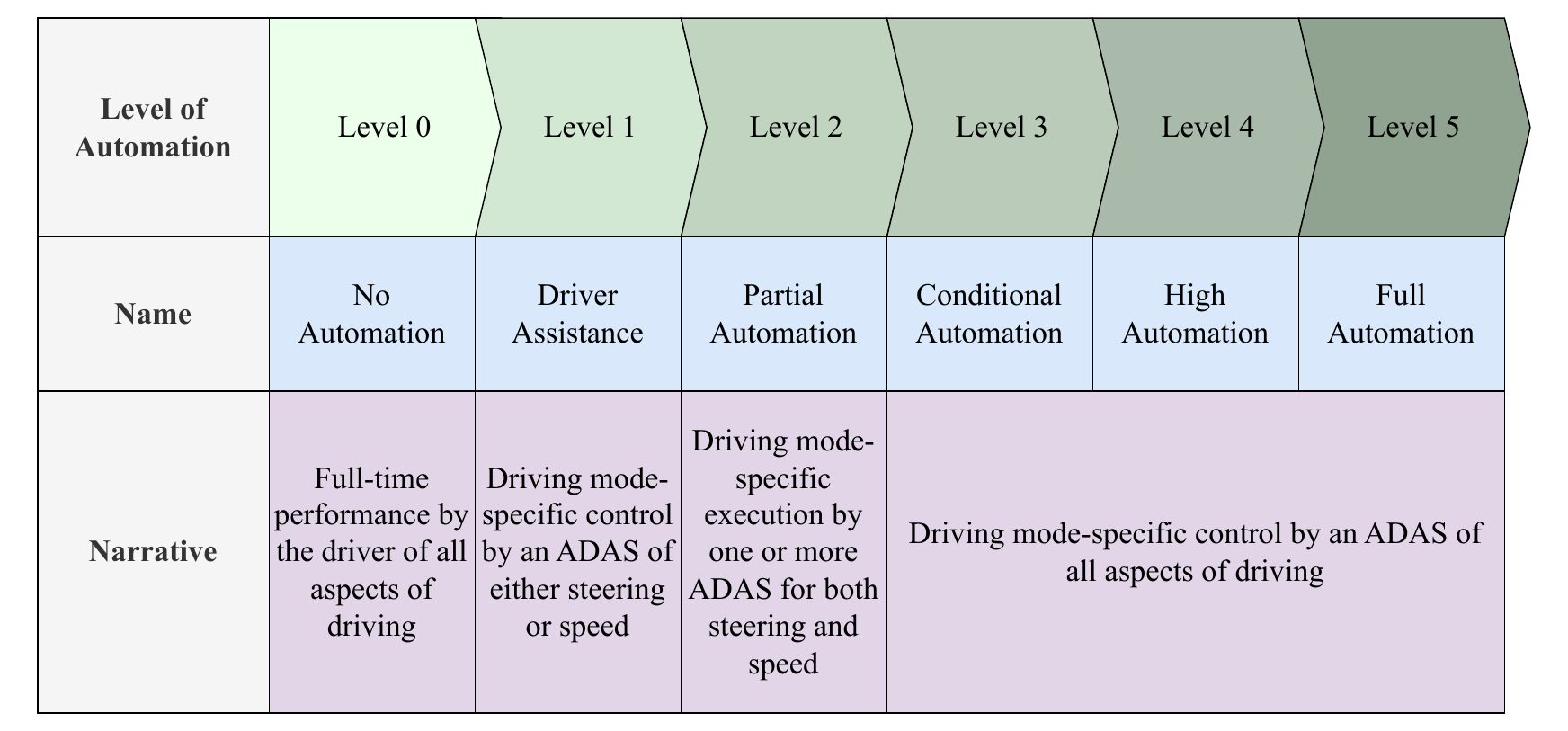}
    \caption{Level of Autonomation of AVs according to the SAE taxonomy}
    \label{fig:level_of_automation}
\end{figure}

\begin{figure}[htbp]
    \centering
    \includegraphics[width=\linewidth]{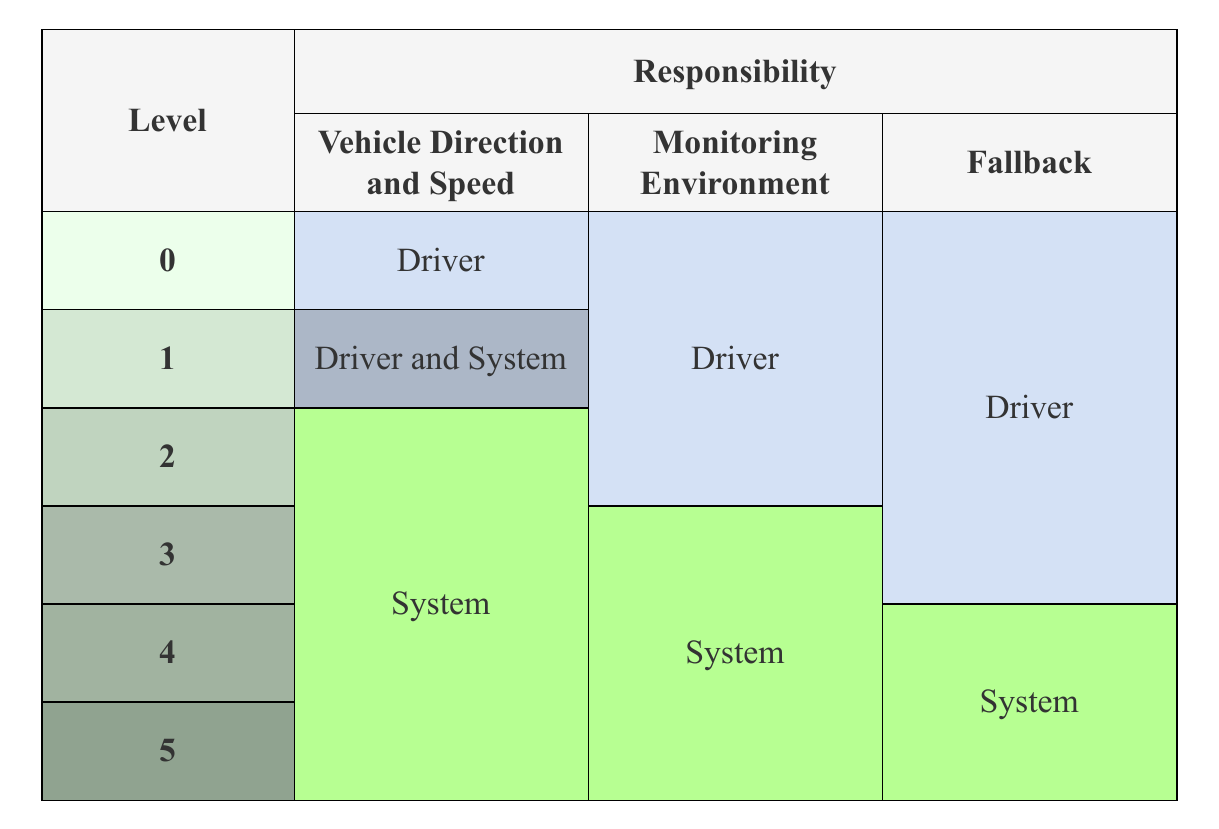}
    \caption{Responsibilities of drivers vs system in different levels of automation}
    \label{fig:av_level_responsibility}
\end{figure}

The development of AVs is often described using the Society of Automotive Engineers (SAE) taxonomy, which defines six levels of automation from Level 0 (no automation) to Level 5 (full automation) \cite{on2021taxonomy}. These levels are briefly described at Figure \ref{fig:level_of_automation}. At Level 1, vehicles support drivers with functions such as adaptive cruise control, while Level 2 combines features like lane-keeping with driver oversight. Level 3 introduces conditional automation, where the system manages most driving tasks but may require human intervention under certain conditions. Levels 4 and 5 represent high and full automation, respectively. Level 4 AVs can operate autonomously within geo-fenced areas or specific scenarios, while Level 5 AVs are envisioned to drive independently in all environments without human input \cite{shladover2018connected}. This stratification highlights the gradual integration of automation technologies and the increasing reliance on complex software systems as autonomous levels advance. Figure \ref{fig:av_level_responsibility} depicts how the responsibilities change between the driver and the automation system across different levels. 

Most recent vehicles of the current time are certified as \textit{Partial} or \textit{Conditional Automation} (level 2 or 3). As of April 2024, only Honda and Mercedes have sold or leased Level 3 cars \cite{Honda2020, Mercedes2024, FORTUNE2024}. Waymo offers Level 4 robotaxi services in parts of Arizona and California without safety drivers \cite{WaymoSite, Reuters2020, kusano2024comparison}. Recent research and development are introducing near-future and experimental products. However, the ultimate goal for autonomous driving is to progress from the existing solutions of level 2 to 4, towards achieving Level 5 Full Automation.

\begin{figure}[htbp]
    \centering
    \includegraphics[width=\linewidth]{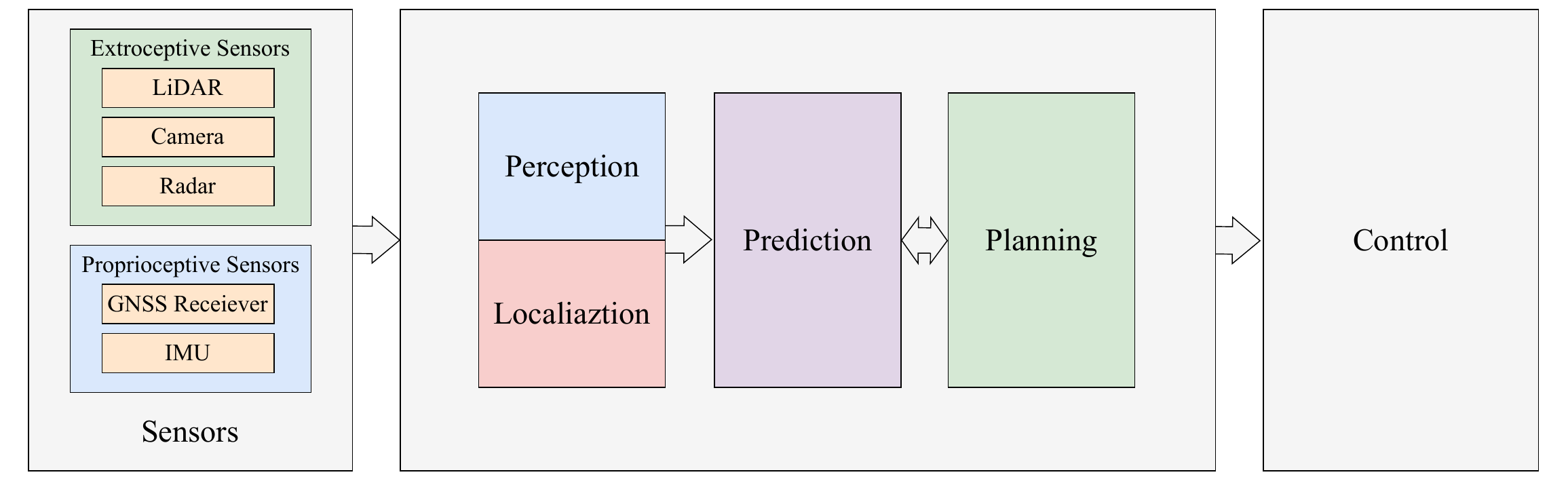}
    \caption{Example general architecture of an AV automation components}
    \label{fig:av_high_level}
\end{figure}

\begin{figure*}[htbp]
    \centering
    \includegraphics[width=.85\linewidth]{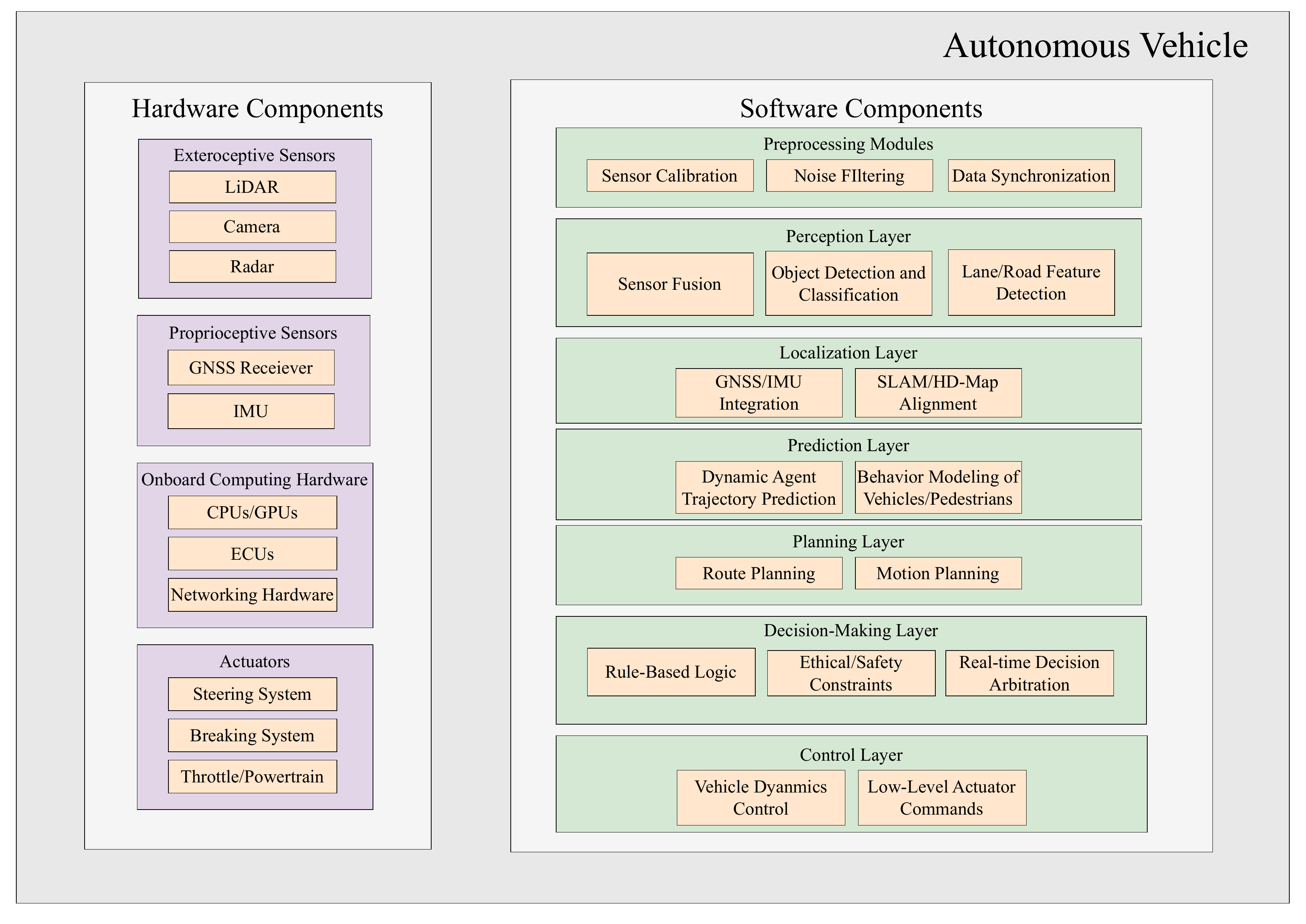}
    \caption{Example hardware and software components of AVs}
    \label{fig:av_components}
\end{figure*}

The architecture of an AV is a tightly coupled ecosystem of hardware and software, designed for safety-critical decision-making in dynamic and uncertain environments. A high-level flow of the core hardware and software modules is illustrated in Figure \ref{fig:av_high_level}. Hardware components include exteroceptive sensors (cameras, LiDAR, radar, ultrasonic sensors) for perception, and proprioceptive sensors (inertial measurement units, GPS, odometry) for localization and motion tracking \cite{ulbrich2017towards}. These components feed data into a layered software stack encompassing perception, localization, prediction, planning, and control modules \cite{yurtsever2020survey}. The decision-making framework integrates outputs from these modules to balance safety, efficiency, and legal/ethical considerations in real time, a safety-critical function even amid uncertain and dynamic environments \cite{liu2021decision, pendleton2017perception}. Figure \ref{fig:av_components} illustrates examples of common hardware and software components of an AV. 

AVs are inherently Software Defined Vehicles (SDVs). An SDV is a vehicle whose core functionality, performance, and user experience are primarily controlled and upgraded through software rather than fixed mechanical or hardware systems. Unlike conventional vehicles, where hardware changes dictate capability, SDVs decouple functionality from hardware by leveraging centralized computing platforms, over-the-air (OTA) updates, and modular software architecture. Perception, decision-making, path planning, and control functionalities of the AVs are entirely orchestrated through complex software stacks that interpret multi-sensor inputs and generate safety-critical actuation commands. This integration makes the software stack the operational backbone of the AVs \cite{teixeira2024deterministic}. The advantages of SDVs include significant flexibility, seamless distribution of new features, improved resource utilization, and support for a wide ecosystem of vehicular services \cite{liu2022impact}. However, this reliance on software also exposes vehicles to heightened cybersecurity and privacy concerns, including over-the-air vulnerabilities, third-party supply-chain risks, and data inference threats, necessitating robust multi-layered defense strategies. System safety, quality assurance, and resilience to software and hardware anomalies are paramount to ensuring functional correctness and preventing catastrophic failure modes \cite{de2024contextualizing}.

The operation of AVs relies on a tightly coupled interplay between hardware and software stacks, where neither can function effectively in isolation. Hardware components, including exteroceptive sensors such as LiDAR, radar, and cameras, along with proprioceptive units like inertial measurement units (IMUs), and GNSS receivers, generate high-volume, heterogeneous data streams that must be fused, interpreted, and acted upon by the vehicle's software stack \cite{grigorescu2020survey}. The data-driven nature of decision-making requires these raw sensor inputs to undergo multi-layered software processing: perception algorithms convert sensor data into scene representation; prediction modules estimate the trajectories of surrounding agents; and planning frameworks translate this contextual understanding into safe, executable motion strategies \cite{pendleton2017perception}.

This interdependency means that data originating from one hardware subsystem is often indispensable for the correct functioning of another. For instance, LiDAR point clouds may be fused with camera imagery through software pipelines to enhance object classification accuracy, which in turn informs motion planning and low-level control executed via actuator hardware \cite{wang2019multi}. A smooth flow of information between sensing hardware and computational modules is thus fundamental to safe and efficient AV behavior.

Critically, it can be stated that sufficient hardware alone is inadequate to ensure safety. Even state-of-the-art sensors are limited by noise, occlusion, adverse weather, and bandwidth constraints. Without robust software to filter out errors, compensate for uncertainties, and enable redundancy, safety cannot be comprehensively achieved \cite{grigorescu2020survey}.

\section{Software Supply Chain of Autonomous Vehicles}
Since the AVs are a type of SDV, the software supply chain in AVs encompasses the full spectrum of activities involved in the development, integration, distribution, and maintenance of vehicle software systems. This supply chain can be broadly described as: (i)upstream open-source libraries (e.g., middleware such as ROS 2, open-source perception or control libraries, different packages, etc.), (ii) third-party commercial modules (e.g., mapping services, localization services, etc.), (iii) proprietary in-house codebases developed by manufacturers for planning, decision-making, and system integration, (iv) development and integration pipelines (e.g., CI/CD infrastructure, automated testing, containeraziation platforms), (v) distribution and update mechanisms (e.g., signed binarie distribution, OTA updates, etc.), and (vi) operational services that support runtime (e.g., cloud-based services and simulation, remote monitoring and diagnostics, etc.). In traditional automotive systems, software is confined to limited embedded functions. In contrast, AVs are software-centric systems that require continuous updates, third-party integrations, and cloud-based services for perception, mapping, connectivity, and safety validation. Weaknesses in any stage of this supply chain, whether from insecure code dependencies, compromised libraries, or misconfigured update mechanisms, pose systemic risks to the safety and reliability of the AVs and passenger safety. As a result, secure life-cycle practices, including software provenance, rigorous testing, and continuous monitoring, are critical \cite{ISO21434}. Organizations such as the National Institute of Standards and Technology (NIST) and the European Union Agency for Cybersecurity (ENISA) highlight the software supply chain assurance critically \cite{ENISA2021, NIST2022}. Considering all these, the life-cycle of AV software can be summarized as: (i) requirements and hazard analysis, (ii) secure design and coding, (iii) integration of third-party components, (iv) testing and verification, (v) signed builds and distribution, and (vi) deployment with post-deployment monitoring and patching. Ensuring integrity across the life cycle requires provenance tracking (e.g., Software Bill of Materials (SBOM)), secure build pipelines, and rigorous update mechanisms.

Autonomous driving stacks exist in both open-source and closed-source forms. Open-source repositories like Autoware, Apollo, and openpilot make their codebases publicly accessible, which brings advantages such as transparency, peer review, faster debugging, reproducibility for academic studies, and flexibility for integration with diverse hardware platforms \cite{kato2018autoware, raju2019performance, chen2022level}. These advantages make these platforms valuable for academic research and early pilot deployments. However, open-source reliance comes with many challenges: security vulnerabilities are publicly visible, maintenance might depend on small developer communities, and performance tuning in real-time environments can be inconsistent. In contrast, closed platforms used by companies like Waymo or Tesla develop their stacks with large in-house engineering teams and controlled hardware environments. This model ensures tighter integration, dedicated security resources, and industrial-scale quality assurance, but at the expense of transparency and community evaluation. Thus, both open and closed source approaches highlight the trade-offs between innovation, transparency, and operational assurance in the AV software supply chain.

The most widely studied open-source platforms, Autoware, Apollo, and openpilot, highlight the diversity of software supply chain design. Autoware is the first widely adopted "all-in-one" open-source AV stack, implemented on ROS/ROS2. It contains modular components for perception, localization, planning, and control, and has been validated in real-vehicle demonstrations, including embedded implementations \cite{kato2018autoware}. Apollo, released by Baidu, is oriented towards production-scale deployment and integrates perception with deep-learning models, high-definition mapping, and planning frameworks. A key distinction is its Cyber RT middleware, which replaces ROS and provides a static scheduling model to improve real-time performance. The Apollo Cyber RT framework is the world’s first open-source, high-performance runtime framework designed specifically for the development of autonomous driving technologies. It was introduced for the first time with the release of the Apollo 3.5 open source platform \cite{belluardo2021multi, ApolloCyberRT}. openpilot has a much narrower scope, limited to Level 2 driver assistance (adaptive cruise control, lane keeping, driver monitoring), but demonstrates how lightweight open-source systems can support consumer use cases \cite{saez2025design}. Comparative studies show that middleware choice can strongly affect performance: for example, ROS 2/DDS as used in Autoware suffers from variable latency under heavy load, while Apollo's Cyber RT achieves lower latency and more deterministic throughput \cite{andreigavrilov17analysis}. Technical comparison between the three platforms is summarized in Table \ref{tab:av_software_comparison}. These architectural differences demonstrate why supply chain robustness depends not only on code provenance but also on middleware performance and update infrastructure, both of which are integral to safe and scalable AV deployment.


\begin{table*}[h]
\centering
\footnotesize
\caption{Technical Comparison of Autoware, Apollo, and openpilot}
\label{tab:av_software_comparison}
\renewcommand{\arraystretch}{1.3}
\begin{tabularx}{\textwidth}{p{3.6cm} X X X}
\toprule
\textbf{Feature} & \textbf{Autoware} & \textbf{Apollo} & \textbf{openpilot} \\
\midrule
\textbf{Primary Language} & C++ and Python & C++ and Python & C++ and Python \\
\textbf{Target Autonomy Level} & L3 -- L4 & L3 -- L4 & L2 \\
\textbf{Middleware} & ROS / ROS 2 (DDS-based, pub-sub) & Cyber RT (custom, static scheduling, high-throughput) & Lightweight in-house messaging framework \\
\textbf{Core Software Modules} & Perception, Localization, Planning, Control & Perception (DL models), Localization, Prediction, Planning, Control, Simulation, HD Maps & Lane Keeping Assist, Adaptive Cruise Control, Driver Monitoring \\
\textbf{Performance Considerations} & ROS latency under load; DDS QoS tuning critical \cite{ye2023ros2} & Cyber RT deterministic scheduling, higher throughput than ROS \cite{andreigavrilov17analysis} & Lightweight but limited scalability beyond L2 \\
\textbf{Community and Ecosystem} & Open-source, large academic and research community & Open-source, strong industry backing (Baidu), contributors from partners & Open-source but maintained primarily by comma.ai \\
\bottomrule
\end{tabularx}
\end{table*}

\section{Autonomous Vehicles Software Supply Chain Vulnerabilities}
Across modern software stacks, supply-chain risks dominate because attackers can compromise systems before code even runs in production, i.e., through malicious packages, typosquatting, dependency confusion, compromised maintainers, poisoned build or CI pipelines, or tampered distribution artifacts. A comprehensive systemization by Ladisa et al. cataloged 107 attack vectors across the contribution, build, and distribution stages and linked them to 94 real incidents, portraying that these threats are not tied to any one language or ecosystem \cite{ladisa2023sok}. These ``SoK" (Systematization of Knowledge) studies and follow-ups show that practical defenses must combine provenance, anomaly detection for packages, and build-time hardening, rather than only relying on a single safeguard \cite{ohm2020towards}.

Open-source AV stacks (e.g., Autoware, Apollo, openpilot) are primarily written in C/C++ (performance-critical perception, planning, and control) plus Python (tooling, glue code, ML workflows). Language-specific risks compound general supply-chain issues, which are discussed in Table \ref{tab:av_vulnerabilities}. In Python, large-scale empirical analyses of the PyPI ecosystem identified 4,669 malicious package files, with ~75\% of the flagged packages reaching end-user projects (often through pip source installs), and frequent behaviors such as credential exfiltration and command execution can clear routes to compromise AV developer environments and CI \cite{guo2023empirical}. In C/C++, peer-reviewed studies and surveys continue to find memory-safety flaws (buffer overflows, use-after-free (UAF), integer overflows, etc.) to be prevalent and exploitable despite mitigation efforts \cite{butt2022depth}; recent evaluations document how attackers still bypass defenses or trigger UAFs that enable control-flow hijack \cite{zhao2024fuzzer}. Together, these findings imply that UAFs that AV codebases inherit both ecosystem-level (package/mirror) and language-level (memory-unsafety) risks, even when upstream modules are popular or long-standing.

For AVs, the quintessential cyber-physical systems, software defects or compromised dependencies can propagate into unsafe vehicle behavior, such as corrupt perception or localization pipelines misclassify/lose obstacles, poisoned models or ROS/ROS 2 message paths desynchronize modules, and runtime memory errors in C/C++ planning/control code can induce denial-of-service or arbitrary actuation. Security studies document exploitable weakness in ROS/ROS 2 isolation and multi-tenant settings \cite{teixeira2020security, xia2025investigating}. Automotive cybersecurity reviews show how attacks on sensors and perception cascade into hazardous decisions, bridging software vulnerabilities and physical safety harms \cite{islam2023review, durlik2024cybersecurity}. Consequently, mitigation is safety-relevant: AV teams must combine supply-chain controls with language-aware hardening and middleware-level protections to prevent defects from becoming hazardous vehicle actions. Table \ref{tab:av_vulnerabilities} illustrates the impact of these vulnerabilities and possible mitigation strategies.

\begin{table*}[h]
\centering
\footnotesize
\caption{Common Software Supply Chain Vulnerabilities in AVs: Impacts and Mitigations}
\label{tab:av_vulnerabilities}
\renewcommand{\arraystretch}{1.3}
\begin{tabularx}{\textwidth}{p{2.8cm} p{2.8cm} p{2.8cm} p{2.8cm} X}
\toprule
\textbf{Vulnerability Type} & \textbf{Example / Source} & \textbf{Impact on AV Stack} & \textbf{Safety Implications} & \textbf{Mitigation Strategies} \\
\midrule
Supply-chain attacks (agnostic) & Dependency confusion, typosquatting & Malicious code injected into OSS modules or CI/CD & Unsafe updates, backdoored perception/control code & Provenance checks, SBOMs, reproducible builds \\
Python package threats & Malicious PyPI packages & Compromised perception/ML preprocessing pipelines & Data poisoning, system hijack & Package vetting, signed package registries \\
C/C++ memory-safety flaws & Buffer overflows, UAFs & Exploits in planning/control loops & DoS, arbitrary actuation commands & Fuzzing, sanitizers, memory-safe rewrites (Rust modules) \\
Middleware weaknesses & ROS 2 isolation flaws & Inter-process message tampering, desync & Loss of perception-control synchronization & DDS security, isolation, QoS tuning \\
Sensor/perception attacks & Spoofing or jamming & Corrupted perception $\rightarrow$ wrong planning inputs & Pedestrian misdetection, unsafe maneuvers & Sensor fusion, anomaly detection, redundancy \\
\bottomrule
\end{tabularx}
\end{table*}

Vulnerabilities are particularly critical in safety-critical systems such as AVs, whether they originate in hardware or software. Since this chapter emphasizes the software supply chain, it is essential to recognize that analyzing vulnerabilities at the system level provides a foundation for mitigating many of these risks. In real-world deployments, compromised software components can propagate through the vehicle stack, leading to unsafe behaviors, accidents, and ultimately threats to human life. As AVs evolve into SDV, systematic risk and threat analysis becomes indispensable to ensure the integrity and resilience of their software supply chain. The following sections, therefore, examine existing analysis methods, modeling tools, and standard approaches that can be applied to vulnerability assessment in the AV context.

\subsection{Existing Vulnerability Analysis Methods}
Software risk or threat analysis systematically identifies plausible attack scenarios, estimates the likelihood or impact, and prioritizes mitigation across the software life cycle. Core steps include- (i) asset or dependency identification, (ii) threat elicitation, (iii) vulnerability enumeration, (iv) risk estimation, and (v) control selection, as formalized in NIST SP 800-30 and widely reused in safety-critical domains. In practice, organizations blend threat-modeling methods with risk analysis: STRIDE (property-centric enumeration of Spoofing, Tampering, Repudiation, Information disclosure, Denial of service, Elevation of privilege), PASTA (Process for Attack Simulation and Threat Analysis; seven-stage, risk-centric "attacker perspective"), LINDDUN (Linking, Identifying, Non-repudiation, Detecting, Data Disclosure, Unawareness, and Non-compliance; privacy-threat focus), attack trees/graphs, and SPTA-Sec-style (Systems-Theoretic Process Analysis for Security) system-hazard analyses for Cyber-Physical Systems (CPS). Systematic reviews document method families, steps, and comparative coverage of security properties \cite{lowe2009stride, xiong2019threat, wolf2021pasta}. These methods complement supply-chain assurance by surfacing risks introduced by third-party libraries, build/CI infrastructure, middleware, and OTA channels \cite{ross2012guide}.

In addition to model-based and process-oriented methods, static analysis tools are widely used to automatically detect code-level weaknesses by scanning source code against known patterns of vulnerabilities. For AV software stacks implemented largely in C/C++ and Python, static analyzers can be applied to the open-source repositories to identify Common Weakness Enumerations (CWEs). Empirical studies demonstrate that static analysis can effectively surface categories of flaws such as buffer overflows, use-after-free, or input-validation errors, which are particularly relevant in performance and safety-critical C/C++ modules \cite{beller2016analyzing, ayewah2010google, livshits2005finding}. While static analysis can suffer from false positives and may not capture runtime or configuration-level threats, it provides an essential first layer of vulnerability analysis in safety-critical domains. Widely used analyzers in this context include Clang Static Analyzer, Cppcheck, Coverty, and CodeQL for C/C++, as well as Bandit and PyLint for Python \cite{beller2016analyzing}. Complementing these, dynamic analysis tools such as Valgrind, AddressSanitizer, and AFL (American Fuzzy Loop) provide runtime monitoring and fuzzing-based testing to uncover memory errors, resource leaks, and input-handling vulnerabilities that a static analyzer may be unable to identify \cite{serebryany2012addresssanitizer, lattner2004llvm, zalewski2017american}. Although for this book chapter, we limited our vulnerability analyses to the static analyzers. 

\subsection{Threat Modeling Tools}
Several tools operationalize the methods above with varying depth and automation. Model-based attack-graph simulators such as securiCAD generate attack paths over formalized meta-models to estimate probabilities, time-to-compromise, and control effects; peer-reviewed studies show their use for enterprise and vehicle domains \cite{mao2019conceptual, xiong2019threat, katsikeas2024empirical}. Strengths include explicit path analysis and "what-if" control evaluation; weaknesses include modeling cost, assumptions about attacker capabilities, and sensitivity to parameterization. Surveyed practice also highlights diagram-centric tools aligned to STRIDE, or PASTA, or LINDDUN that facilitate data-flow modeling and catalog-based threat enumeration. Literatures report strong onboarding and communication benefits, but limited quantification and coverage bias without domain-specific tailoring \cite{xiong2019threat}. On the other hand, static analyzers are not threat-modeling tools, but the vulnerability data they produce, e.g., CWE mappings, can be integrated into higher-level threat models or attack graphs. For AV stacks, tools that (i) capture supply-chain structure (dependencies, build provenance), (ii) reflect middleware/runtime interactions (ROS 2/DDS, Cyber RT), and (iii) support attack-graph reasoning are most suitable for analyzing cascading safety effects from software compromise.

\subsection{Standard Approaches}
Standardized approaches provide the foundation for assessing risks in any system by offering structured, repeatable processes. In the context of AVs, these approaches are essential for aligning general software assurance practices with domain-specific safety and cybersecurity requirements. This section outlines both general frameworks and automotive-specific practices, before considering their adaptation to the AV software stacks:

\subsubsection{General Software Analysis Standards}
The NIST Secure Software Development Framework (SSDF) prescribes organization and build-level practices (e.g., hardened builds, code provenance, review/testing, vulnerability response) and explicitly maps to federal acquisition expectations. It is framework-agnostic and intended to overlay any Software Development Life Cycle (SDLC) \cite{souppaya2022secure}. For risk assessment, NIST SP 800-30 remains the canonical method (risk framing, assessment, response, monitoring) and is often combined with control catalogs or domain guidance \cite{ross2012guide}. SBOM standards underpin supply-chain transparency by tracking upstream components and transitive dependencies \cite{xia2023empirical}.

\subsubsection{Autonomous Vehicle Specific Practices}
In road vehicles, ISO/SAE 21434 mandates TARA (Threat Analysis and Risk Assessment) integrated across concept, development, production, and post-production; UN R155 further requires a Cybersecurity Management System (CSMS) for type approval, operationalizing risk-based processes over the vehicle life cycle. Academic evaluations propose and compare automotive TARA methods, e.g., HEAVENS 2.0 aligns with ISO/SAE 21434 while addressing gaps (asset/threat cataloging, attack-feasibility, risk scoring) and has seen industrial uptake; case studies combine TARA with STRIDE for connected vehicle functions \cite{lautenbach2021proposing, boi2023strengthening}. These standards/frameworks primarily target system-level cyber-risk and interface threats; they do not fully standardize software supply-chain assurance, necessitating adoption of SSDF or SBOM controls alongside ISO, SAE, and UN requirements for AV software.

\subsubsection{Adapting to AV Software Stacks}
For perception-localization-planning-control pipelines, combining automotive TARA (safety system context) with software supply-chain controls (SSDF tasks; SBOM for dependency provenance) produces traceability from third-party components to safety goals \cite{NIST2022, xia2023empirical}. The advantages are end-to-end coverage from upstream code through middleware/runtime, measurable build attestations, and repeatable patch/recall decisions, although it can cause modeling overhead.

While standards such as ISO/SAE 21434, UN R155, and NIST SSDF provide structured processes for conducting vulnerability and risk assessments, they do not in themselves resolve the question of how to prioritize risks across the software supply chain. Static analyzers, SBOM-based provenance tracking, and model-driven assessments surface numerous weaknesses that are often mapped to CWEs; organizations require a systematic way to compare and rank these findings according to both technical severity and safety impact. This gap is addressed through risk scoring frameworks, which translate identified vulnerabilities into actionable metrics that guide mitigation priorities.

Risk can be scored qualitatively for speed and stakeholder alignment, or quantitatively to express loss exposure and support cost-benefit trade-offs. Common Vulnerability Scoring System (CVSS) v4.0 standardizes vulnerability severity at the component level with Base / Threat / Environment / Supplemental metrics. It is useful for triaging AV stack dependencies, but does not capture system-level safety impact by itself \cite{cvss4, aggarwal2023study}. On the other hand, Factor Analysis of Information Risk (FAIR) style quantitative risk analysis models the frequency and magnitude of loss. Peer-reviewed work shows Bayesian/FAIR approaches can support monetary risk decisions and investment optimization, complementing qualitative TARA outputs \cite{wang2020bayesian, he2024cybersecurity}.

\section{Case Study: Vulnerability Analyses of Autoware, Apollo, and openpilot Platforms}
In this section, we present an approach for vulnerability analysis for three different open-source AV software packages (i.e., Autoware, Apollo, and openpilot) and corresponding findings related to code and third-party library vulnerabilities.

\subsection{Approach for Vulnerability Analyses}
\begin{figure}[htbp]
    \centering
    \includegraphics[width=.8\linewidth]{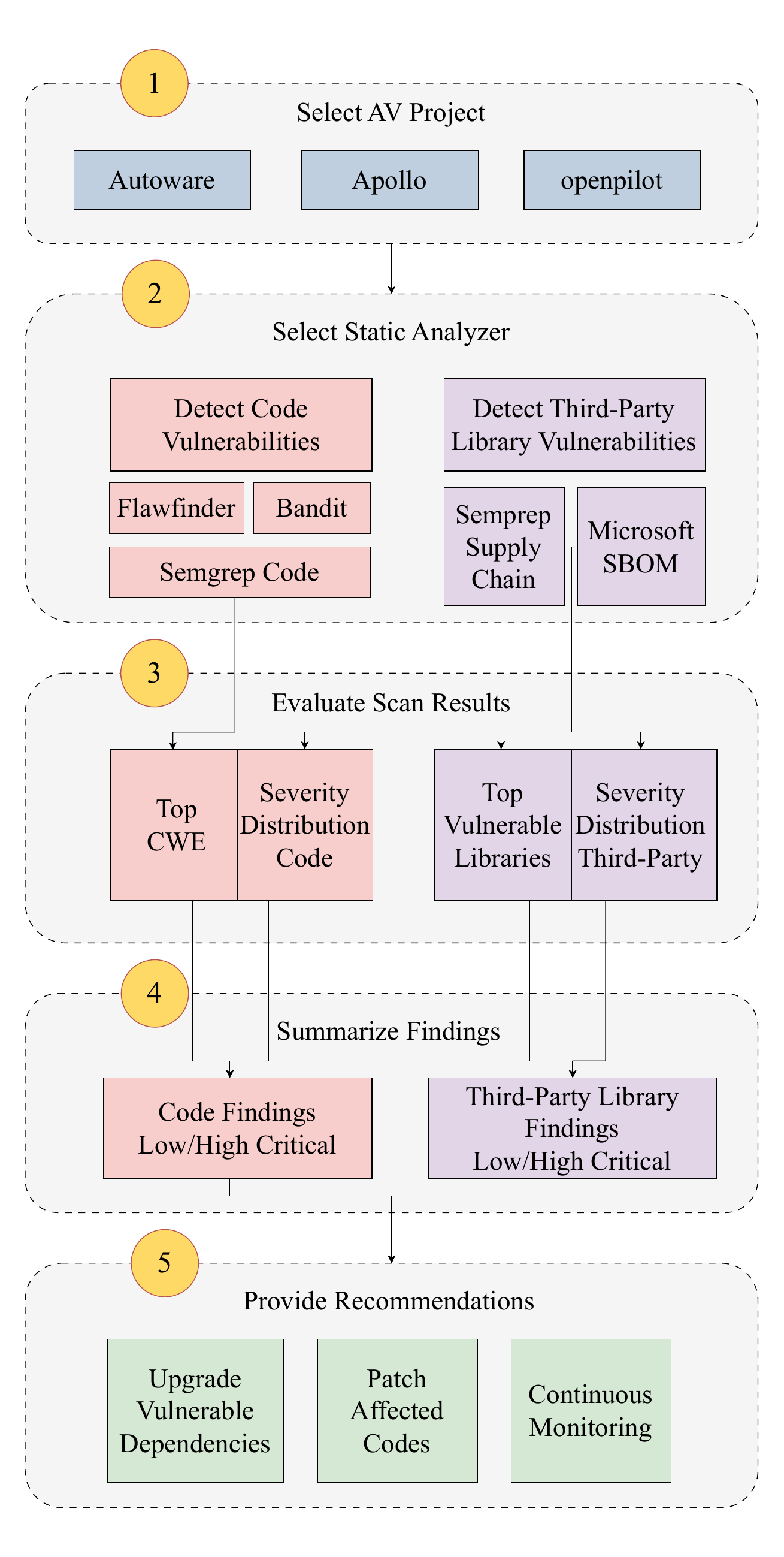}
    \caption{Vulnerability analysis workflow}
    \label{fig:method_flowchart}
\end{figure}

The approach for analyzing vulnerabilities in AV software stack consists of a set of steps, as illustrated in Figure~\ref{fig:method_flowchart}. The workflow follows a structured, stepwise approach from repository selection to actionable recommendations. The first step involves selecting the AV software repositories for analysis. In this case study, we select three open-source repositories: openpilot, Autoware, and Apollo. These repositories provide a diverse set of Python and C/C++ codebases representative of common AV software components. Once projects are selected, static analysis is conducted to detect potential vulnerabilities in the code and associated dependencies. This phase has two main components: (i) code vulnerability identification, and (ii) third-party library vulnerability identification. Tools, such as Flawfinder (for C/C++), Bandit (for Python) and Semgrep, are employed to identify coding vulnerabilities, including buffer overflows, command injections, insecure input handling, and other critical weaknesses. Dependencies and third-party packages are analyzed using Semgrep’s supply chain rules and Microsoft SBOM tools to detect outdated, insecure, or vulnerable components. After analyzing, the results are evaluated to understand the severity and distribution of vulnerabilities. The findings are categorized by Common Weakness Enumerations (CWEs), and severity distributions are analyzed to identify the most critical issues. Dependency vulnerabilities are prioritized based on severity and reachability, highlighting the most at-risk components.  The critical and high-severity vulnerabilities from both code and third-party library analyses are summarized. This provides a comprehensive overview of the major risks present across the selected AV repositories and facilitates comparison between different types of vulnerabilities. Finally, actionable recommendations are provided to mitigate the identified risks, including upgrading vulnerable dependencies to patched versions, patching affected code to resolve critical and high-severity issues, and continuous monitoring of codebases and dependencies to proactively address future vulnerabilities.

\subsection{Source Code Findings}
In this subsection, we applied multiple static analysis tools, i.e., Flawfinder, Bandit, and Semgrep, across Apollo, Autoware, and openpilot. Using a combination of language-specific and cross-language analyzers enabled us to capture a wide range of vulnerabilities, from low-level memory errors in C/C++ to insecure coding practices in Python. These tools allowed us to identify recurring weaknesses, map them to CWE categories, and assess their severity to better understand the security posture of each repository. This comparative analysis not only reveals repository-specific risks but also uncovers broader vulnerability patterns, offering valuable insights for AV researchers and practitioners to strengthen security during development and deployment.

\subsubsection{Flawfinder and Bandit Analysis Overview}
Static analysis tools like Bandit\cite{banditTool}, and Flawfinder\cite{wheelerFlawfinder} provide developers with automated ways to identify potential security vulnerabilities in their code before deployment. Bandit focuses on Python, scanning source files to identify insecure coding patterns, cryptographic weaknesses, and hardcoded secrets, while Flawfinder analyzes C/C++ code to detect common risks, such as buffer overflows, format string issues, race conditions, and unsafe system calls. Both tools generate prioritized reports that help developers focus on the most critical security issues, though they are not substitutes for secure coding knowledge or human review. In Table \ref{tab:flawfinder_bandit_summary}, we present a summary of the Flawfinder and Bandit analyses conducted on three open-source AV repositories: Apollo, Autoware\_core, and Openpilot. Table \ref{tab:flawfinder_bandit_summary} captures both C/C++ and Python files, showing the total number of analyzed files, statistical metrics (mean, median, maximum, and minimum) for the number of lines per vulnerable file, and the total number of identified vulnerabilities for each repository and tool. This summary provides a quick overview of the scale of analysis and the distribution of vulnerabilities across different languages and repositories.


\begin{table*}[htbp]
\centering
\caption{Summary of Flawfinder and Bandit Analysis}
\label{tab:flawfinder_bandit_summary}
\resizebox{\linewidth}{!}{%
\begin{tabular}{l l r r r r r r r}
\toprule
\multicolumn{2}{c}{} & \multicolumn{1}{c}{\textbf{Total Analyzed Files}} & \multicolumn{4}{c}{\textbf{Number of Lines Per Vulnerable Files}} & \multicolumn{1}{c}{\textbf{Total Vulnerable Files}} & \multicolumn{1}{c}{\textbf{Total CWEs}} \\
\cmidrule(lr){3-3} \cmidrule(lr){4-7} \cmidrule(lr){8-8} \cmidrule(lr){9-9}
\textbf{Project} & \textbf{Tool} & & \textbf{Mean} & \textbf{Median} & \textbf{Max} & \textbf{Min} & & \\
\midrule
Apollo        & C/C++ files   & 1957 & 1471.52 & 991.5 & 4523  & 11 & 424  & 604  \\
Apollo        & Python files  & 314  & 182.50  & 92    & 956   & 5  & 121  & 121  \\
Autoware\_core & C/C++ files  & 605  & 154.44  & 155   & 593   & 19 & 25   & 26   \\
Autoware\_core & Python files & 25   & 78.44   & 111   & 128   & 18 & 9    & 9    \\
Openpilot     & C/C++ files   & 432  & 2906.45 & 329   & 16827 & 10 & 227  & 310  \\
Openpilot     & Python files  & 481  & 123.43  & 93    & 808   & 1  & 1179 & 1179 \\
\bottomrule
\end{tabular}%
}
\end{table*}

From Table \ref{tab:flawfinder_bandit_summary}, we made several observations. Firstly, Openpilot C/C++ files have the highest mean and maximum number of lines per vulnerable file, indicating that some files are very large and potentially complex. Apollo Python files show a moderate number of vulnerabilities with a relatively low mean and median number of lines, suggesting smaller but frequent issues. Autoware-core Python files have a very small number of analyzed files and vulnerabilities, but the median and maximum line counts indicate that a few files contain larger segments of vulnerable code. Overall, the total vulnerabilities are highest in openpilot Python files, reflecting the intensive analysis of Python code in that repository. The statistical spread between mean, median, min, and max provides insight into variability, showing that some repositories have large outliers in file size while others are more consistent.

\subsubsection{Analysis of Common Weakness Enumerations (CWEs)}
\begin{figure*}[h]
    \centering
    \includegraphics[width=\linewidth]{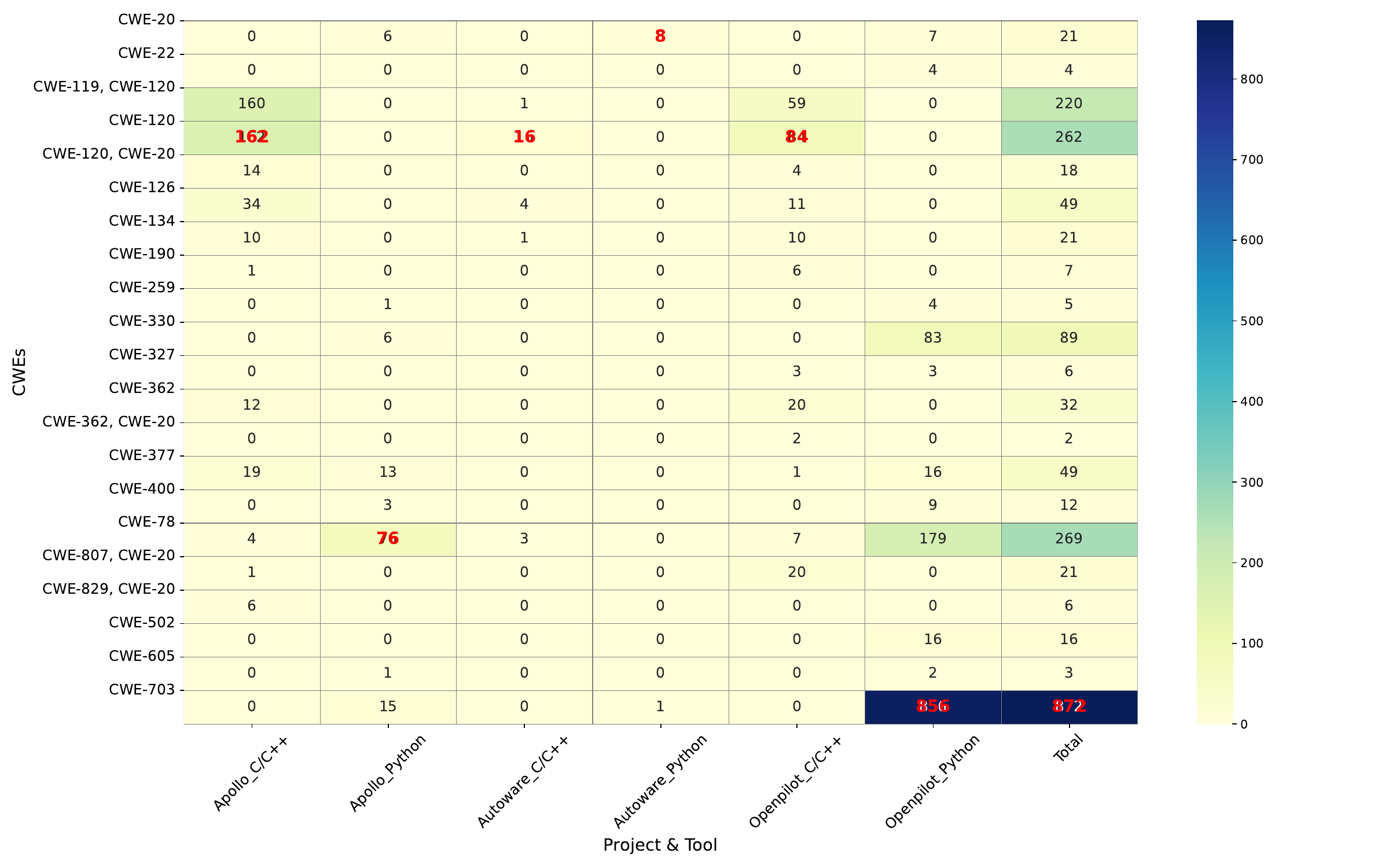}
    \caption{Distribution of CWE Issues Detected Across Projects and Tools}
    \label{fig:issue_distribution}
\end{figure*}

Figure \ref{fig:issue_distribution} presents a detailed view of the CWE issues identified across three AV open-source software packages: Apollo, Autoware\_core, and openpilot. Both C/C++ and Python files were analyzed using Flawfinder and Bandit, respectively. These data capture the distribution of vulnerabilities across different CWE types for various AV repositories and programming languages. The heatmap provides a visual representation of the relative frequency of each CWE, with the maximum occurrences highlighted to facilitate quick insights into the most prevalent vulnerabilities.

Table \ref{tab:cwe_list} lists the selected CWEs that were most frequently observed across the analyzed AV repositories. Each CWE is accompanied by a brief description and notes on its potential impact, providing readers with a quick reference to understand the nature of common vulnerabilities in both C/C++ and Python code. Among the most frequent issues across all repositories are CWE-703, CWE-120, and CWE-119/CWE-120, indicating recurring patterns of vulnerabilities in both C/C++ and Python code. Examining repository-specific trends, Openpilot Python files exhibit the highest total number of vulnerabilities, largely driven by CWE-703, which reflects the intensive use of Python code with security-relevant issues. In contrast, Apollo C/C++ files show a substantial number of CWE-119 and CWE-120 vulnerabilities, suggesting that certain legacy C/C++ modules may require closer review. Tool-specific patterns also become evident: some CWEs, such as CWE-330 and CWE-362, are primarily detected in Python files, highlighting the importance of using Bandit alongside Flawfinder to cover language-specific security weaknesses. Finally, distribution insights from the heatmap reveal both the concentration and spread of vulnerabilities, with red-highlighted maximum values indicating the CWE-repository combinations that are most critical, thereby helping prioritize remediation efforts.

\begin{table*}[htbp]
\centering
\footnotesize
\caption{Selected Common Weakness Enumerations (CWEs)}
\label{tab:cwe_list}
\renewcommand{\arraystretch}{1.2}
\begin{tabularx}{\textwidth}{p{2.5cm} p{5cm} X}
\toprule
\textbf{CWE ID} & \textbf{Name} & \textbf{Impact / Notes} \\
\midrule
CWE-20 \cite{cwe20}  & Improper Input Validation & Root cause of many attacks (injection, deserialization, etc.) \\
CWE-22 \cite{cwe22}  & Path Traversal & Arbitrary file access possible \\
CWE-78 \cite{cwe78}  & OS Command Injection & Remote code execution risk \\
CWE-119 \cite{cwe119} & Improper Restriction of Operations within Memory Buffer & Parent category of buffer overflows \\
CWE-120 \cite{cwe120} & Buffer Copy without Checking Size of Input & Classic buffer overflow \\
CWE-126 \cite{cwe126} & Buffer Over-read & Can cause leaks of sensitive info (e.g., Heartbleed bug) \\
CWE-134 \cite{cwe134} & Use of Externally-Controlled Format String & Can lead to code execution / memory corruption \\
CWE-190 \cite{cwe190} & Integer Overflow or Wraparound & May cause memory corruption or bypass checks \\
CWE-259 \cite{cwe259} & Use of Hard-coded Password & Credential disclosure risk \\
CWE-327 \cite{cwe327} & Use of Broken or Risky Cryptographic Algorithm & Weak crypto (e.g., MD5, DES) \\
CWE-330 \cite{cwe330} & Use of Insufficiently Random Values & Predictable keys/tokens \\
CWE-362 \cite{cwe362} & Race Condition & Can cause data corruption or privilege escalation \\
CWE-377 \cite{cwe377} & Insecure Temporary File & Exploitable in multi-user environments \\
CWE-400 \cite{cwe400} & Uncontrolled Resource Consumption (DoS) & Denial-of-service risk \\
CWE-502 \cite{cwe502} & Deserialization of Untrusted Data & Remote code execution \\
CWE-605 \cite{cwe605} & Multiple Binds to the Same Port & Service hijacking or denial of service \\
CWE-703 \cite{cwe703} & Improper Check or Handling of Exceptional Conditions & Error handling flaws leading to bypasses \\
CWE-807 \cite{cwe807} & Reliance on Untrusted Inputs in Security Decision & May allow privilege escalation \\
CWE-829 \cite{cwe829} & Inclusion of Functionality from Untrusted Control Sphere & Attacker-controlled code execution \\
\bottomrule
\end{tabularx}
\end{table*}

\subsection{Top 10 CWE Issues Across AV repositories}
To better visualize the distribution of the most prevalent vulnerabilities across the three AV open-source repositories, i.e., Apollo, Autoware\_core, and Openpilot, a pie chart was generated for the top 10 CWE issues. The chart summarizes the total occurrences of each CWE type, combining both C/C++ and Python analyses from Flawfinder and Bandit. This graphical representation helps quickly identify which vulnerabilities are most common and potentially critical, providing an at-a-glance understanding of the security landscape across these repositories.

\begin{figure}[h]
    \centering
    \includegraphics[width=\linewidth]{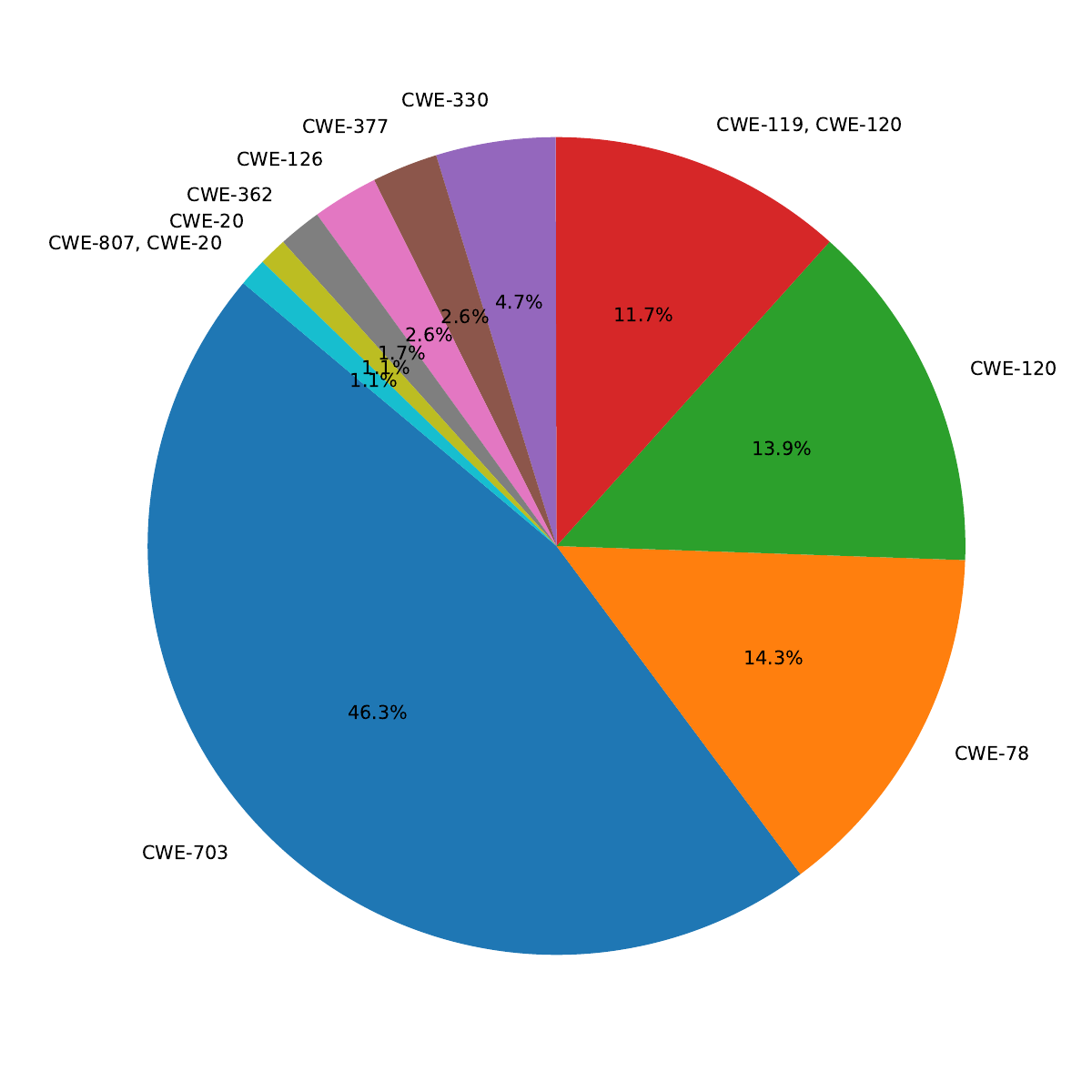}
    \caption{Top-10 CWE Issues Across All Projects}
    \label{fig:pie_chart}
\end{figure}

The pie chart highlights several key insights from the top 10 CWEs across three AV repositories. CWE-703 dominates the distribution, largely driven by Openpilot Python files, indicating frequent use of insecure code patterns that could affect runtime security. Following this, CWE-120 and CWE-119/CWE-120 are the next most prevalent issues, primarily observed in C/C++ files, pointing to potential buffer overflow and memory management risks in legacy code. Other CWEs, such as CWE-330, CWE-78, and CWE-377, occur moderately across three AV repositories, reflecting recurring weaknesses that affect both Python and C/C++ components. Overall, the chart provides a visual ranking of vulnerabilities, enabling one to quickly identify the most critical weaknesses that require attention in these AV repositories.

\subsubsection{Semgrep Code Finding Analysis}
Semgrep\cite{semgrep} is a lightweight, open-source static analysis tool that helps developers detect bugs, enforce coding standards, and uncover security vulnerabilities. Unlike simple text-based search tools, Semgrep understands code structure and semantics, making it effective for scanning across more than 30 programming languages. Its rule-based approach allows flexible customization, making it widely used for secure software development. In our analysis, we applied Semgrep to all three AV repositories. The tool reported both code findings (direct vulnerabilities in the source code) and third-party library findings (issues originating from external dependencies). The results reveal varying risk profiles across AV repositories, with Apollo showing the highest number of both code and third-party library issues, while Autoware Core had comparatively fewer findings. The Semgrep results indicate that Apollo has the largest security exposure, with 71 code findings and 91 third-party library issues across a wide dependency base of over 1,500 libraries. Openpilot shows a notable number of 57 code findings, but only one third-party issue, suggesting its risks are more localized within its own source code. In contrast, Autoware Core has very limited findings, with just seven code-related issues and no third-party library vulnerabilities, which may reflect a smaller or more contained codebase.


\begin{table}[t] 
\centering
\caption{Semgrep Static Analysis Findings Across Projects}
\label{tab:semgrep_findings}
\begin{tabularx}{\linewidth}{l *{3}{>{\centering\arraybackslash}X}}
\toprule
\textbf{Project} & \textbf{Code Findings} & \textbf{Supply Chain Findings} & \textbf{Scanned Dependencies} \\
\midrule
Apollo        & 71 & 91 & 1{,}515 \\
Openpilot     & 57 & 1  & 314     \\
Autoware Core & 7  & 0  & 0       \\
\bottomrule
\end{tabularx}
\end{table}

\subsubsection{Semgrep Findings}

To better understand the security weaknesses identified by Semgrep, the code findings were mapped to their corresponding CWE identifiers. Table~\ref{tab:semgrep_code_cwe_8col} presents this mapping, showing the repository, CWE, severity, and the number of code findings for each issue. This layout highlights both the type and severity of vulnerabilities across the three AV repositories, as well as the frequency of their occurrence, allowing one to quickly identify the most prevalent and critical security issues.

\begin{table*}[htbp]
\centering
\caption{Semgrep Coding Findings Mapped to CWEs (Two-Side Layout)}
\label{tab:semgrep_code_cwe_8col}
\renewcommand{\arraystretch}{1.1} 
\small
\resizebox{\linewidth}{!}{%
\begin{tabularx}{\linewidth}{l l l X l l l X}
\toprule
\textbf{Project} & \textbf{CWE} & \textbf{Severity} & \textbf{Code Findings} & 
\textbf{Project} & \textbf{CWE} & \textbf{Severity} & \textbf{Code Findings} \\
\midrule
Autoware  & CWE-78   & High   & 4  & Openpilot & CWE-476 & High   & 3  \\
Autoware  & CWE-611  & High   & 2  & Openpilot & CWE-78  & High   & 12 \\
Autoware  & CWE-913  & Medium & 1  & Openpilot & CWE-78  & High   & 3  \\
Openpilot & CWE-427  & High   & 2  & Openpilot & CWE-22  & Medium & 2  \\
Openpilot & CWE-798  & High   & 1  & Openpilot & CWE-611 & High   & 1  \\
Openpilot & CWE-125  & Medium & 1  & Openpilot & CWE-502 & Medium & 9  \\
Openpilot & CWE-353  & Medium & 6  & Openpilot & CWE-706 & Medium & 5  \\
Openpilot & CWE-939  & Medium & 3  & Openpilot & CWE-327 & Medium & 1  \\
Openpilot & CWE-276  & Medium & 1  & Openpilot & CWE-95  & Low    & 1  \\
Openpilot & CWE-78   & Low    & 4  & Openpilot & CWE-319 & Low    & 2  \\
Apollo    & CWE-416  & High   & 2  & Apollo    & CWE-22   & High   & 1  \\
Apollo    & CWE-78   & High   & 7  & Apollo    & CWE-79   & High   & 4  \\
Apollo    & CWE-319  & High   & 4  & Apollo    & CWE-611  & High   & 2  \\
Apollo    & CWE-120  & Medium & 3  & Apollo    & CWE-125  & Medium & 2  \\
Apollo    & CWE-22   & Medium & 13 & Apollo    & CWE-20   & Medium & 9  \\
Apollo    & CWE-676  & Medium & 4  & Apollo    & CWE-116  & Medium & 2  \\
Apollo    & CWE-353  & Medium & 2  & Apollo    & CWE-668  & Medium & 1  \\
Apollo    & CWE-96   & Medium & 1  & Apollo    & CWE-95   & Medium & 1  \\
Apollo    & CWE-706  & Medium & 1  & Apollo    & CWE-1333 & Medium & 1  \\
Apollo    & CWE-78   & Low    & 4  & Apollo    & CWE-134  & Low    & 7  \\
\bottomrule
\end{tabularx}
}
\end{table*}

\subsubsection{Semgrep's Top 10 CWEs}

In order to highlight the most frequent security weaknesses detected by Semgrep, we summarized the top 10 CWE identifiers based on coding findings across the three AV repositories. This table shows both the type of vulnerability and the number of occurrences, providing a clear overview of the most critical coding issues. From Table \ref{tab:top10_semgrep_coding}, several observations can be made. \textbf{CWE-78} (OS Command Injection) is the most frequent issue, indicating that multiple modules are vulnerable to command execution risks. \textbf{CWE-22} (Path Traversal) is the second most common vulnerability, highlighting potential risks related to unauthorized file access. Other vulnerabilities, such as \textbf{CWE-502}, \textbf{CWE-20}, and \textbf{CWE-353}, are also notable, reflecting issues in deserialization, input validation, and improper data handling. Overall, these top 10 CWEs emphasize areas where developers should focus code review and remediation efforts to enhance the security of the AV repositories.

{\setlength{\tabcolsep}{20pt}{

\begin{table}[tb]
\centering
\caption{Top 10 Coding Findings Mapped to CWEs}
\label{tab:top10_semgrep_coding}
\begin{tabular}{c c c}
\toprule
\textbf{S.No} & \textbf{CWE} & \textbf{Coding Findings} \\
\midrule
1  & CWE-78   & 34 \\
2  & CWE-22   & 16 \\
3  & CWE-502  & 9  \\
4  & CWE-20   & 9  \\
5  & CWE-353  & 8  \\
6  & CWE-134  & 7  \\
7  & CWE-706  & 6  \\
8  & CWE-319  & 6  \\
9  & CWE-611  & 5  \\
10 & CWE-79   & 4  \\
\bottomrule
\end{tabular}
\end{table}
}

\subsubsection{CWE and Severity Distribution in Autonomous Vehicle Projects}

To better understand the security issues detected by Semgrep, we plotted pie charts at Figure \ref{fig:cwe_severity_pie} showing the distribution of vulnerabilities by CWE (inner layer) and severity (outer layer) for each AV repository. From the pie charts, we observe that for Autoware, the majority of issues are medium to high severity, mostly related to CWE-78 and CWE-22. Openpilot shows a mix of high and medium severity issues, with CWE-78, CWE-502, and CWE-353 being more frequent. Apollo has several high-severity findings in CWE-78 and CWE-79, but also a considerable number of medium-severity issues. Overall, the charts help to quickly identify which repositories have more critical weaknesses and which CWE categories need more attention for fixes and code review. The severity distribution outer layer is especially useful to spot the critical areas quickly.

\begin{figure*}[htbp]
    \centering
    \includegraphics[width=0.4\linewidth]{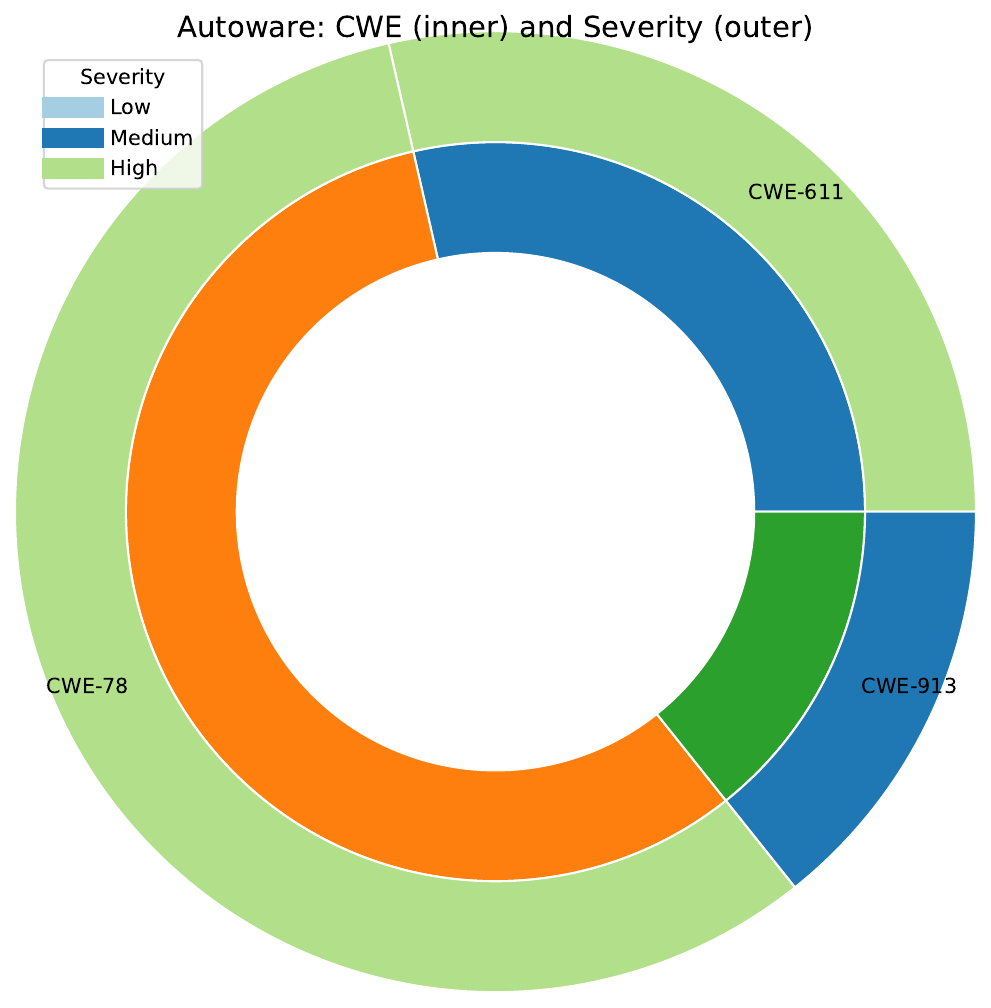}\\[0.1em]
    \includegraphics[width=0.4\linewidth]{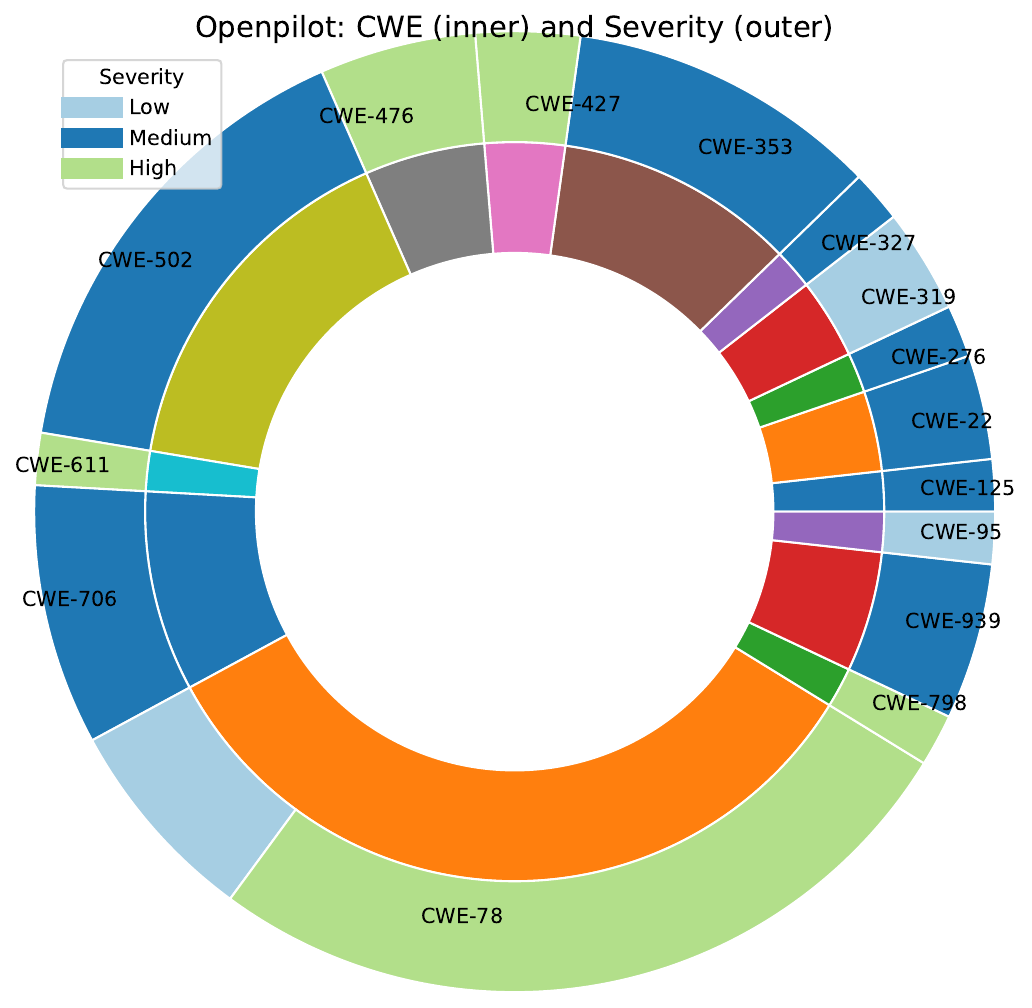}\\[0.1em]
    \includegraphics[width=0.4\linewidth]{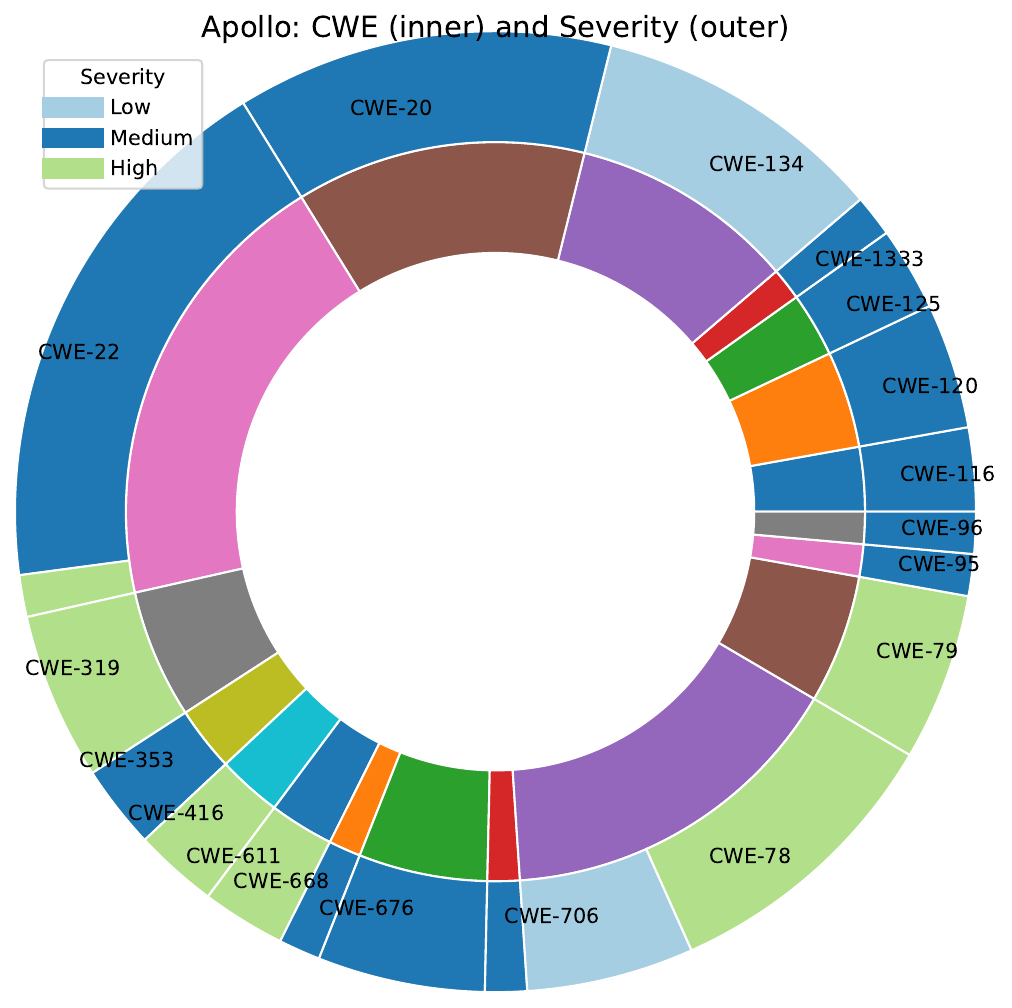}
    \caption{CWE (inner) and Severity (outer) distribution for all AV repositories. Top: Autoware, Middle: Openpilot, Bottom: Apollo.}
    \label{fig:cwe_severity_pie}
\end{figure*}

\subsubsection{Key Findings of Code Vulnerability}
Across all three AV repositories, several CWEs appear repeatedly in both coding findings (Semgrep) and static analysis (Flawfinder/Bandit). CWE-78 (OS Command Injection) and CWE-22 (Path Traversal) are consistently among the top vulnerabilities, indicating risks in both Python and C/C++ code. Other recurring issues include CWE-502, CWE-20, and CWE-353, reflecting common input validation and deserialization weaknesses. Buffer and memory-related CWEs, such as CWE-119, CWE-120, and CWE-126, are notable in legacy C/C++ modules. Overall, these recurring CWEs suggest that certain classes of vulnerabilities are pervasive across repositories and languages, and they should be prioritized during code review and remediation.

\subsection{Third-Party Library Vulnerabilities}
Modern AV software stacks make extensive use of third-party libraries and dependencies. While this speeds up development and brings useful functionality, it also creates significant risks in the software supply chain. Even if the main application code is carefully written and reviewed, weaknesses in external components can expose the system to serious threats such as privilege escalation or remote code execution. To understand these risks, we examined the selected AV repositories with Semgrep SBOM scanning and the Microsoft SBOM tool combined with Grype, which allowed us to detect vulnerabilities in dependencies, group them by severity, and evaluate their potential impact on overall system security.

\subsubsection{Semgrap Supply Chain Vulnerability Analysis}
During the Semgrep SBOM scan of the selected repositories, we observed notable differences in dependency vulnerabilities. For the Apollo AV repositories, the scan identified a total of 91 dependencies that require further review. Among these, some dependencies were marked as \textbf{reachable}, indicating that they are actively used in the code and could potentially be exploited, while others were marked as \textbf{need-review}, requiring manual assessment to determine their actual impact. Additionally, several dependencies were \textbf{unreachable}, meaning they are present in the repository but not currently used, though they still pose a potential risk if brought into use in the future. In contrast, the openpilot AV repositories had only 1 reachable dependency, and Autoware\_Core had none. The vulnerabilities in the Apollo AV repositories, categorized by severity, are shown in Table~\ref{tab:severityofapolloproject}. For example, the tar-fs package contains a high-severity vulnerability related to path traversal. The output looks like,

\begin{verbatim}
>>> tar-fs - CVE-2024-12905
    Severity: HIGH                                  
    Affected versions of tar-fs are vulnerable to
    Improper Limitation of a Pathname to a 
    Restricted Directory ('Path Traversal').
    tar-fs improperly resolves 
    symlinks and fails to restrict pathnames during
    archive extraction. This 
    vulnerability can allow an attacker
    to craft a tar archive that writes or overwrites
    files outside the intended directory, potentially
    compromising the host system.       
    
    Fixed for tar-fs at versions: 1.16.4, 2.1.2, 
    3.0.7
    10516| tar-fs@^2.0.0, tar-fs@^2.1.1:
\end{verbatim}

{\setlength{\tabcolsep}{50pt}


\begin{table}[h]
\centering
\footnotesize
\caption{Vulnerabilities Categorized by Severity for Apollo Project}
\label{tab:severityofapolloproject}
\begin{tabular}{l c}
\toprule
\textbf{Severity} & \textbf{Count} \\
\midrule
Critical & 15 \\
High     & 37 \\
Medium   & 29 \\
Low      & 10 \\
\bottomrule
\end{tabular}
\end{table}

}

Affected versions of tar-fs fail to properly restrict pathnames during archive extraction, allowing attackers to overwrite files outside the intended directory. This issue has been fixed in versions 1.16.4, 2.1.2, and 3.0.7. Based on the Semgrep SBOM scan of the Apollo AV repositories, dependencies flagged as critical must be addressed immediately by upgrading to the latest secure versions, as they represent the highest potential risk if exploited. High-severity dependencies could be prioritized for short-term mitigation or upgrades, while medium- and low-severity dependencies could be monitored and updated regularly to prevent cumulative security exposure. Reachable dependencies require particular attention since they are actively used in the codebase, whereas dependencies marked as needing review should be assessed manually to determine their exploitability. Even unreachable dependencies could be periodically reviewed to ensure that future use does not introduce vulnerabilities. Overall, maintaining a routine of SBOM scanning and timely dependency updates is essential to minimize risk and ensure that both direct and transitive dependencies remain secure.

\subsubsection{Microsoft SBOM Analysis}
We performed an SBOM analysis on multiple open-source repositories using Microsoft's sbom-tool\cite{microsoft_sbom_tool} and vulnerability scanning with Grype\cite{anchore_grype}. The goal was to identify known vulnerabilities in third-party components and dependencies within these repositories. For each repository, the workflow involved cloning the repository, generating the SBOM using sbom-tool, and scanning it with Grype. The sbom-tool parameters included specifying the build path (-bc), output folder (-b), package name (-pn), version (-pv), and namespace URL (-nsb). For example, the result generation process for the SBOM tool for the Apollo repository is,

\begin{verbatim}
sbom-tool generate \
  -b ./sbom-output \
  -bc . \
  -pn "apollo" \
  -pv "1.0.0" \
  -ps "ApolloAuto" \
  -nsb "https://github.com/ApolloAuto/apollo"
\end{verbatim}

For Autoware\_Core, the SBOM analysis did not detect any vulnerabilities, suggesting that the dependencies in this repository are either up-to-date or do not have known security issues listed in the vulnerability databases. In the case of openpilot, the scan detected 3 vulnerabilities, with severity distribution: 1 high and 2 medium. Notable components included \texttt{yajl-ruby} and \texttt{rake} as provided in Table \ref{tab:OpenPilotVulnerabilitiesDetectedfromSBOMScan}. The SBOM scan for Apollo identified 1,532 components, of which 84 had known vulnerabilities. The severity distribution included 10 critical, 35 high, 29 medium, 10 low, and 0 negligible vulnerabilities. Selected critical and high-severity components are summarized in Table \ref{tab:SelectedCriticalandHighVulnerabilitiesinApollofromSBOMScan}. The analysis suggests that immediate attention should be given to critical vulnerabilities, such as those found in \texttt{ejs} and \texttt{loader-utils}, by applying patches or upgrading to secure versions. High-severity vulnerabilities, including multiple versions of \texttt{json5}, could be addressed in the short term to minimize potential exploitation. Furthermore, it is recommended to perform regular SBOM scans and maintain up-to-date dependencies to continuously reduce security risks and ensure the overall integrity of the software repositories.


\begin{table*}[h]
\centering
\footnotesize
\caption{Openpilot Vulnerabilities Detected from SBOM Scan}
\label{tab:OpenPilotVulnerabilitiesDetectedfromSBOMScan}
\resizebox{\linewidth}{!}{%
\begin{tabular}{l l l l l l l l}
\toprule
\textbf{Component} & \textbf{Installed} & \textbf{Fixed In} & \textbf{Type} & \textbf{Vulnerability} & \textbf{Severity} & \textbf{EPSS} & \textbf{Risk} \\
\midrule
yajl-ruby & 1.2.1 & 1.3.1 & gem & GHSA-wwh7-4jw9-33x6 & High   & 1.0\%  & 0.7   \\
yajl-ruby & 1.2.1 & 1.4.3 & gem & GHSA-jj47-x69x-mxrm & Medium & 1.1\%  & 0.6   \\
rake      & 10.3.2 & 12.3.3 & gem & GHSA-jppv-gw3r-w3q8 & Medium & $<0.1$\% & $<0.1$ \\
\bottomrule
\end{tabular}
}
\end{table*}


\begin{table*}[htbp]
\centering
\caption{Selected Critical and High Vulnerabilities in Apollo from SBOM Scan}
\label{tab:SelectedCriticalandHighVulnerabilitiesinApollofromSBOMScan}
\resizebox{\linewidth}{!}{%
\begin{tabular}{l l l l l l l l}
\toprule
\textbf{Name} & \textbf{Installed} & \textbf{Fixed In} & \textbf{Type} & \textbf{Vulnerability} & \textbf{Severity} & \textbf{EPSS} & \textbf{Risk} \\
\midrule
ejs           & 2.7.4  & 3.1.7  & npm & GHSA-phwq-j96m-2c2q & Critical & 93.5\% (99th) & 87.9 \\
json5         & 0.5.1  & 1.0.2  & npm & GHSA-9c47-m6qq-7p4h & High     & 40.5\% (97th) & 29.6 \\
json5         & 1.0.1  & 1.0.2  & npm & GHSA-9c47-m6qq-7p4h & High     & 40.5\% (97th) & 29.6 \\
json5         & 2.2.1  & 2.2.2  & npm & GHSA-9c47-m6qq-7p4h & High     & 40.5\% (97th) & 29.6 \\
loader-utils  & 0.2.17 & 1.4.1  & npm & GHSA-76p3-8jx3-jpfq & Critical & 15.8\% (94th) & 14.9 \\
loader-utils  & 1.4.0  & 1.4.1  & npm & GHSA-76p3-8jx3-jpfq & Critical & 15.8\% (94th) & 14.9 \\
loader-utils  & 2.0.2  & 2.0.3  & npm & GHSA-76p3-8jx3-jpfq & Critical & 15.8\% (94th) & 14.9 \\
\bottomrule
\end{tabular}
}
\end{table*}

\subsubsection{Key Findings of Third-Party Library Vulnerabilities}The analysis of the selected repositories using Semgrep and Microsoft SBOM tools revealed notable third-party library vulnerabilities across multiple dependencies. Several dependencies were flagged as reachable, needing review, or unreachable, with issues spanning Critical, High, Medium, and Low severity levels. Critical and high-severity vulnerabilities were primarily associated with outdated or insecure package versions, such as tar-fs, node-forge, and postcss, potentially enabling attacks like path traversal, prototype pollution, or cryptographic signature forgery. These results emphasize the importance of proactive dependency management, timely patching, and continuous monitoring to reduce security risks in the software supply chain.

\section{Threats to Validity}
While our study provides insights into code and third-party library vulnerabilities in AV open-source repositories, several factors could influence the validity and generalizability of our findings, such as, (i) internal validity; (ii) external validity; (iii) construct validity; and (iv) reliability and reproducibility. Internal validity concerns the accuracy of our vulnerability detection. Although we used well-established tools, such as Flawfinder, Bandit, Semgrep, Microsoft SBOM tool, and Grype, no automated tool guarantees a comprehensive vulnerability identification. False positives and false negatives may exist, potentially skewing our quantitative findings. Additionally, mapping code findings to CWE identifiers may introduce minor misclassifications due to differing tool interpretations. Our study focuses on three open-source AV repositories: Apollo, Openpilot, and Autoware. While these AV repositories are representative of widely used autonomous software, the results may not generalize to other autonomous systems or repositories in different programming languages. Moreover, dependency and supply chain risks observed here might not apply to other environments with different libraries, versions, or configurations. Construct validity relates to whether our measures accurately capture the security properties of interest. CWEs, severity levels, and dependency classifications provide standardized measures, but they may not fully represent runtime exploitation risks. Our experiments depend on specific repository versions and tool versions.  Reproducing the study exactly requires access to the same code snapshots and tool configurations. The datasets, analysis scripts, and supplementary materials used in this study are publicly available on GitHub: \href{https://github.com/uaerfan/cav-security-vulnerabilities.git}{https://github.com/uaerfan/cav-security-vulnerabilities.git}. Although we provide the GitHub link and Docker images, scripts to facilitate replication, differences in environments, future dependency updates, or changes in tool rules may lead to slightly different results.

\section{Conclusion}
This study analyzed the software supply chain of AVs through static vulnerability assessment of three widely used open-source platforms- Autoware, Apollo, and openpilot, revealing recurring CWEs and dependency-related risks that highlight the safety-critical nature of securing AV software ecosystems. Our findings underscore that while static analyzers such as Flawfinder and Bandit provide meaningful insights into prevalent vulnerabilities, they also illustrate that open-source AV software stacks remain exposed to systematic supply-chain threats that must be addressed proactively. However, the study is constrained by practical limitations: (i) only C/C++ and Python files were analyzed, leaving other language components unexamined, (ii) static analysis alone cannot capture dynamic or runtime-only vulnerabilities, and (iii) the results reflect a snapshot of the repositories at a fixed point in time, which may not represent evolving vulnerability landscapes. These limitations motivate future research directions: expanding analyses to cover additional programming languages and middleware, integrating dynamic analysis and fuzzing techniques to capture runtime risks, and adopting continuous monitoring pipelines to track vulnerability changes across evolving repositories. Addressing these gaps will advance a more comprehensive, real-time vulnerability management framework for AV software supply chains, thereby strengthening resilience and ensuring safety in increasingly software-defined vehicles.
 
Despite these threats and limitations, the study provides meaningful insights into prevalent vulnerabilities and supply chain risks, highlighting areas for remediation and further research.

\bibliographystyle{ACM-Reference-Format}
\bibliography{citations}


\begin{thebibliography}{85}


\ifx \showCODEN    \undefined \def \showCODEN     #1{\unskip}     \fi
\ifx \showISBNx    \undefined \def \showISBNx     #1{\unskip}     \fi
\ifx \showISBNxiii \undefined \def \showISBNxiii  #1{\unskip}     \fi
\ifx \showISSN     \undefined \def \showISSN      #1{\unskip}     \fi
\ifx \showLCCN     \undefined \def \showLCCN      #1{\unskip}     \fi
\ifx \shownote     \undefined \def \shownote      #1{#1}          \fi
\ifx \showarticletitle \undefined \def \showarticletitle #1{#1}   \fi
\ifx \showURL      \undefined \def \showURL       {\relax}        \fi
\providecommand\bibfield[2]{#2}
\providecommand\bibinfo[2]{#2}
\providecommand\natexlab[1]{#1}
\providecommand\showeprint[2][]{arXiv:#2}

\bibitem[Aggarwal(2023)]%
        {aggarwal2023study}
\bibfield{author}{\bibinfo{person}{Manuj Aggarwal}.} \bibinfo{year}{2023}\natexlab{}.
\newblock \showarticletitle{A study of CVSS v4. 0: A CVE scoring system}. In \bibinfo{booktitle}{\emph{2023 6th International Conference on Contemporary Computing and Informatics (IC3I)}}, Vol.~\bibinfo{volume}{6}. IEEE, \bibinfo{pages}{1180--1186}.
\newblock


\bibitem[Aliane(2025)]%
        {aliane2025survey}
\bibfield{author}{\bibinfo{person}{Nourdine Aliane}.} \bibinfo{year}{2025}\natexlab{}.
\newblock \showarticletitle{A Survey of Open-Source Autonomous Driving Systems and Their Impact on Research}.
\newblock \bibinfo{journal}{\emph{Information}} \bibinfo{volume}{16}, \bibinfo{number}{4} (\bibinfo{year}{2025}), \bibinfo{pages}{317}.
\newblock


\bibitem[Anchore(2025)]%
        {anchore_grype}
\bibfield{author}{\bibinfo{person}{Inc. Anchore}.} \bibinfo{year}{2025}\natexlab{}.
\newblock \bibinfo{title}{Grype: A Vulnerability Scanner for Container Images and Filesystems}.
\newblock
\urldef\tempurl%
\url{https://github.com/anchore/grype}
\showURL{%
\tempurl}
\newblock
\shownote{Accessed: 2025-09-18}.


\bibitem[Anderson et~al\mbox{.}(2014)]%
        {anderson2014autonomous}
\bibfield{author}{\bibinfo{person}{James~M Anderson}, \bibinfo{person}{Kalra Nidhi}, \bibinfo{person}{Karlyn~D Stanley}, \bibinfo{person}{Paul Sorensen}, \bibinfo{person}{Constantine Samaras}, {and} \bibinfo{person}{Oluwatobi~A Oluwatola}.} \bibinfo{year}{2014}\natexlab{}.
\newblock \bibinfo{booktitle}{\emph{Autonomous vehicle technology: A guide for policymakers}}.
\newblock \bibinfo{publisher}{Rand Corporation}.
\newblock


\bibitem[AndreiGavrilov et~al\mbox{.}({[n.\,d.]})]%
        {andreigavrilov17analysis}
\bibfield{author}{\bibinfo{person}{Marlen~Bergaliyev AndreiGavrilov}, \bibinfo{person}{Sergey Tinyakov}, {and} \bibinfo{person}{Kirill Krinkin}.} \bibinfo{year}{[n.\,d.]}\natexlab{}.
\newblock \showarticletitle{Analysis Of Robotic Platforms: Data Transfer Performance Evaluation}.
\newblock \bibinfo{journal}{\emph{architecture}}  \bibinfo{volume}{17} (\bibinfo{year}{[n.\,d.]}), \bibinfo{pages}{18}.
\newblock


\bibitem[Ayewah and Pugh(2010)]%
        {ayewah2010google}
\bibfield{author}{\bibinfo{person}{Nathaniel Ayewah} {and} \bibinfo{person}{William Pugh}.} \bibinfo{year}{2010}\natexlab{}.
\newblock \showarticletitle{The google findbugs fixit}. In \bibinfo{booktitle}{\emph{Proceedings of the 19th international symposium on Software testing and analysis}}. \bibinfo{pages}{241--252}.
\newblock


\bibitem[Beller et~al\mbox{.}(2016)]%
        {beller2016analyzing}
\bibfield{author}{\bibinfo{person}{Moritz Beller}, \bibinfo{person}{Radjino Bholanath}, \bibinfo{person}{Shane McIntosh}, {and} \bibinfo{person}{Andy Zaidman}.} \bibinfo{year}{2016}\natexlab{}.
\newblock \showarticletitle{Analyzing the state of static analysis: A large-scale evaluation in open source software}. In \bibinfo{booktitle}{\emph{2016 IEEE 23rd International Conference on Software Analysis, Evolution, and Reengineering (SANER)}}, Vol.~\bibinfo{volume}{1}. IEEE, \bibinfo{pages}{470--481}.
\newblock


\bibitem[Belluardo et~al\mbox{.}(2021)]%
        {belluardo2021multi}
\bibfield{author}{\bibinfo{person}{Luca Belluardo}, \bibinfo{person}{Andrea Stevanato}, \bibinfo{person}{Daniel Casini}, \bibinfo{person}{Giorgiomaria Cicero}, \bibinfo{person}{Alessandro Biondi}, {and} \bibinfo{person}{Giorgio Buttazzo}.} \bibinfo{year}{2021}\natexlab{}.
\newblock \showarticletitle{A multi-domain software architecture for safe and secure autonomous driving}. In \bibinfo{booktitle}{\emph{2021 IEEE 27th international conference on embedded and real-time computing systems and applications (RTCSA)}}. IEEE, \bibinfo{pages}{73--82}.
\newblock


\bibitem[Boi et~al\mbox{.}(2023)]%
        {boi2023strengthening}
\bibfield{author}{\bibinfo{person}{Biagio Boi}, \bibinfo{person}{Tarush Gupta}, \bibinfo{person}{Marcelo Rinhel}, \bibinfo{person}{Iuliana Jubea}, \bibinfo{person}{Rahamatullah Khondoker}, \bibinfo{person}{Christian Esposito}, {and} \bibinfo{person}{Bruno~Miguel Sousa}.} \bibinfo{year}{2023}\natexlab{}.
\newblock \showarticletitle{Strengthening automotive cybersecurity: a comparative analysis of ISO/SAE 21434-compliant automatic collision notification (ACN) systems}.
\newblock \bibinfo{journal}{\emph{Vehicles}} \bibinfo{volume}{5}, \bibinfo{number}{4} (\bibinfo{year}{2023}), \bibinfo{pages}{1760--1802}.
\newblock


\bibitem[Butt et~al\mbox{.}(2022)]%
        {butt2022depth}
\bibfield{author}{\bibinfo{person}{Muhammad~Arif Butt}, \bibinfo{person}{Zarafshan Ajmal}, \bibinfo{person}{Zafar~Iqbal Khan}, \bibinfo{person}{Muhammad Idrees}, {and} \bibinfo{person}{Yasir Javed}.} \bibinfo{year}{2022}\natexlab{}.
\newblock \showarticletitle{An in-depth survey of bypassing buffer overflow mitigation techniques}.
\newblock \bibinfo{journal}{\emph{Applied Sciences}} \bibinfo{volume}{12}, \bibinfo{number}{13} (\bibinfo{year}{2022}), \bibinfo{pages}{6702}.
\newblock


\bibitem[Chen et~al\mbox{.}(2022)]%
        {chen2022level}
\bibfield{author}{\bibinfo{person}{Li Chen}, \bibinfo{person}{Tutian Tang}, \bibinfo{person}{Zhitian Cai}, \bibinfo{person}{Yang Li}, \bibinfo{person}{Penghao Wu}, \bibinfo{person}{Hongyang Li}, \bibinfo{person}{Jianping Shi}, \bibinfo{person}{Junchi Yan}, {and} \bibinfo{person}{Yu Qiao}.} \bibinfo{year}{2022}\natexlab{}.
\newblock \showarticletitle{Level 2 autonomous driving on a single device: Diving into the devils of openpilot}.
\newblock \bibinfo{journal}{\emph{arXiv preprint arXiv:2206.08176}} (\bibinfo{year}{2022}).
\newblock


\bibitem[Co.(2020)]%
        {Honda2020}
\bibfield{author}{\bibinfo{person}{Honda~Motor Co.}} \bibinfo{year}{2020}\natexlab{}.
\newblock \bibinfo{booktitle}{\emph{Honda Sensing Elite with Traffic Jam Pilot: World’s first approval of Level 3 automated driving system}}.
\newblock Honda.
\newblock
\urldef\tempurl%
\url{https://global.honda/en/newsroom/news/2020/4201111eng.html}
\showURL{%
\tempurl}


\bibitem[Committee(2021)]%
        {on2021taxonomy}
\bibfield{author}{\bibinfo{person}{On-Road Automated Driving~(ORAD) Committee}.} \bibinfo{year}{2021}\natexlab{}.
\newblock \bibinfo{booktitle}{\emph{Taxonomy and definitions for terms related to driving automation systems for on-road motor vehicles}}.
\newblock \bibinfo{publisher}{SAE international}.
\newblock


\bibitem[De~Vincenzi et~al\mbox{.}(2024)]%
        {de2024contextualizing}
\bibfield{author}{\bibinfo{person}{Marco De~Vincenzi}, \bibinfo{person}{Mert~D Pes{\'e}}, \bibinfo{person}{Chiara Bodei}, \bibinfo{person}{Ilaria Matteucci}, \bibinfo{person}{Richard~R Brooks}, \bibinfo{person}{Monowar Hasan}, \bibinfo{person}{Andrea Saracino}, \bibinfo{person}{Mohammad Hamad}, {and} \bibinfo{person}{Sebastian Steinhorst}.} \bibinfo{year}{2024}\natexlab{}.
\newblock \showarticletitle{Contextualizing security and privacy of software-defined vehicles: State of the art and industry perspectives}.
\newblock \bibinfo{journal}{\emph{arXiv preprint arXiv:2411.10612}} (\bibinfo{year}{2024}).
\newblock


\bibitem[Durlik et~al\mbox{.}(2024)]%
        {durlik2024cybersecurity}
\bibfield{author}{\bibinfo{person}{Irmina Durlik}, \bibinfo{person}{Tymoteusz Miller}, \bibinfo{person}{Ewelina Kostecka}, \bibinfo{person}{Zenon Zwierzewicz}, {and} \bibinfo{person}{Adrianna {\L}obodzi{\'n}ska}.} \bibinfo{year}{2024}\natexlab{}.
\newblock \showarticletitle{Cybersecurity in autonomous vehicles—are we ready for the challenge?}
\newblock \bibinfo{journal}{\emph{Electronics}} \bibinfo{volume}{13}, \bibinfo{number}{13} (\bibinfo{year}{2024}), \bibinfo{pages}{2654}.
\newblock


\bibitem[{ENISA}(2021)]%
        {ENISA2021}
\bibfield{author}{\bibinfo{person}{{ENISA}}.} \bibinfo{year}{2021}\natexlab{}.
\newblock \bibinfo{booktitle}{\emph{{Threat Landscape for Supply Chain Attacks}}}.
\newblock \bibinfo{type}{{T}echnical {R}eport}. \bibinfo{institution}{European Union Agency for Cybersecurity (ENISA)}, \bibinfo{address}{Heraklion, Greece}.
\newblock


\bibitem[Fagnant and Kockelman(2015)]%
        {fagnant2015preparing}
\bibfield{author}{\bibinfo{person}{Daniel~J Fagnant} {and} \bibinfo{person}{Kara Kockelman}.} \bibinfo{year}{2015}\natexlab{}.
\newblock \showarticletitle{Preparing a nation for autonomous vehicles: opportunities, barriers and policy recommendations}.
\newblock \bibinfo{journal}{\emph{Transportation Research Part A: Policy and Practice}}  \bibinfo{volume}{77} (\bibinfo{year}{2015}), \bibinfo{pages}{167--181}.
\newblock


\bibitem[{FIRST.Org, Inc.}(2023)]%
        {cvss4}
\bibfield{author}{\bibinfo{person}{{FIRST.Org, Inc.}}} \bibinfo{year}{2023}\natexlab{}.
\newblock \bibinfo{booktitle}{\emph{Common Vulnerability Scoring System v4.0: Specification Document}}.
\newblock
\urldef\tempurl%
\url{https://www.first.org/cvss/v4.0/specification-document}
\showURL{%
\tempurl}
\newblock
\shownote{Accessed: 2025-09-17}.


\bibitem[Grigorescu et~al\mbox{.}(2020)]%
        {grigorescu2020survey}
\bibfield{author}{\bibinfo{person}{Sorin Grigorescu}, \bibinfo{person}{Bogdan Trasnea}, \bibinfo{person}{Tiberiu Cocias}, {and} \bibinfo{person}{Gigel Macesanu}.} \bibinfo{year}{2020}\natexlab{}.
\newblock \showarticletitle{A survey of deep learning techniques for autonomous driving}.
\newblock \bibinfo{journal}{\emph{Journal of field robotics}} \bibinfo{volume}{37}, \bibinfo{number}{3} (\bibinfo{year}{2020}), \bibinfo{pages}{362--386}.
\newblock


\bibitem[Group(2024)]%
        {Mercedes2024}
\bibfield{author}{\bibinfo{person}{Mercedes-Benz Group}.} \bibinfo{year}{2024}\natexlab{}.
\newblock \bibinfo{booktitle}{\emph{DRIVE PILOT approved for Level 3 automated driving in California and Nevada}}.
\newblock Mercedes-Benz.
\newblock
\urldef\tempurl%
\url{https://group.mercedes-benz.com/innovation/product-innovation/autonomous-driving/drive-pilot-california.html}
\showURL{%
\tempurl}


\bibitem[Guo et~al\mbox{.}(2023)]%
        {guo2023empirical}
\bibfield{author}{\bibinfo{person}{Wenbo Guo}, \bibinfo{person}{Zhengzi Xu}, \bibinfo{person}{Chengwei Liu}, \bibinfo{person}{Cheng Huang}, \bibinfo{person}{Yong Fang}, {and} \bibinfo{person}{Yang Liu}.} \bibinfo{year}{2023}\natexlab{}.
\newblock \showarticletitle{An empirical study of malicious code in pypi ecosystem}. In \bibinfo{booktitle}{\emph{2023 38th IEEE/ACM International Conference on Automated Software Engineering (ASE)}}. IEEE, \bibinfo{pages}{166--177}.
\newblock


\bibitem[He et~al\mbox{.}(2024)]%
        {he2024cybersecurity}
\bibfield{author}{\bibinfo{person}{Ying He}, \bibinfo{person}{Tong Xin}, {and} \bibinfo{person}{Cunjin Luo}.} \bibinfo{year}{2024}\natexlab{}.
\newblock \showarticletitle{Cybersecurity Investments Metrics using FAIR-ROSI}.
\newblock  (\bibinfo{year}{2024}).
\newblock


\bibitem[Islam et~al\mbox{.}(2023)]%
        {islam2023review}
\bibfield{author}{\bibinfo{person}{Taminul Islam}, \bibinfo{person}{Md~Alif Sheakh}, \bibinfo{person}{Anjuman~Naher Jui}, \bibinfo{person}{Omar Sharif}, {and} \bibinfo{person}{Md~Zobaer Hasan}.} \bibinfo{year}{2023}\natexlab{}.
\newblock \showarticletitle{A review of cyber attacks on sensors and perception systems in autonomous vehicle}.
\newblock \bibinfo{journal}{\emph{Journal of Economy and Technology}}  \bibinfo{volume}{1} (\bibinfo{year}{2023}), \bibinfo{pages}{242--258}.
\newblock


\bibitem[{ISO/SAE}(2021)]%
        {ISO21434}
\bibfield{author}{\bibinfo{person}{{ISO/SAE}}.} \bibinfo{year}{2021}\natexlab{}.
\newblock \bibinfo{booktitle}{\emph{{ISO/SAE 21434:2021 Road Vehicles – Cybersecurity Engineering}}}.
\newblock \bibinfo{type}{Standard} ISO/SAE 21434:2021. \bibinfo{institution}{International Organization for Standardization (ISO) and SAE International}, \bibinfo{address}{Geneva, Switzerland}.
\newblock


\bibitem[Kato et~al\mbox{.}(2018)]%
        {kato2018autoware}
\bibfield{author}{\bibinfo{person}{Shinpei Kato}, \bibinfo{person}{Shota Tokunaga}, \bibinfo{person}{Yuya Maruyama}, \bibinfo{person}{Seiya Maeda}, \bibinfo{person}{Manato Hirabayashi}, \bibinfo{person}{Yuki Kitsukawa}, \bibinfo{person}{Abraham Monrroy}, \bibinfo{person}{Tomohito Ando}, \bibinfo{person}{Yusuke Fujii}, {and} \bibinfo{person}{Takuya Azumi}.} \bibinfo{year}{2018}\natexlab{}.
\newblock \showarticletitle{Autoware on board: Enabling autonomous vehicles with embedded systems}. In \bibinfo{booktitle}{\emph{2018 ACM/IEEE 9th International Conference on Cyber-Physical Systems (ICCPS)}}. IEEE, \bibinfo{pages}{287--296}.
\newblock


\bibitem[Katsikeas et~al\mbox{.}(2024)]%
        {katsikeas2024empirical}
\bibfield{author}{\bibinfo{person}{Sotirios Katsikeas}, \bibinfo{person}{Engla~Rencelj Ling}, \bibinfo{person}{Pontus Johnsson}, {and} \bibinfo{person}{Mathias Ekstedt}.} \bibinfo{year}{2024}\natexlab{}.
\newblock \showarticletitle{Empirical evaluation of a threat modeling language as a cybersecurity assessment tool}.
\newblock \bibinfo{journal}{\emph{Computers \& Security}}  \bibinfo{volume}{140} (\bibinfo{year}{2024}), \bibinfo{pages}{103743}.
\newblock


\bibitem[Kusano et~al\mbox{.}(2024)]%
        {kusano2024comparison}
\bibfield{author}{\bibinfo{person}{Kristofer~D Kusano}, \bibinfo{person}{John~M Scanlon}, \bibinfo{person}{Yin-Hsiu Chen}, \bibinfo{person}{Timothy~L McMurry}, \bibinfo{person}{Ruoshu Chen}, \bibinfo{person}{Tilia Gode}, {and} \bibinfo{person}{Trent Victor}.} \bibinfo{year}{2024}\natexlab{}.
\newblock \showarticletitle{Comparison of Waymo rider-only crash data to human benchmarks at 7.1 million miles}.
\newblock \bibinfo{journal}{\emph{Traffic Injury Prevention}} \bibinfo{volume}{25}, \bibinfo{number}{sup1} (\bibinfo{year}{2024}), \bibinfo{pages}{S66--S77}.
\newblock


\bibitem[Ladisa et~al\mbox{.}(2023)]%
        {ladisa2023sok}
\bibfield{author}{\bibinfo{person}{Piergiorgio Ladisa}, \bibinfo{person}{Henrik Plate}, \bibinfo{person}{Matias Martinez}, {and} \bibinfo{person}{Olivier Barais}.} \bibinfo{year}{2023}\natexlab{}.
\newblock \showarticletitle{Sok: Taxonomy of attacks on open-source software supply chains}. In \bibinfo{booktitle}{\emph{2023 IEEE Symposium on Security and Privacy (SP)}}. IEEE, \bibinfo{pages}{1509--1526}.
\newblock


\bibitem[Lattner and Adve(2004)]%
        {lattner2004llvm}
\bibfield{author}{\bibinfo{person}{Chris Lattner} {and} \bibinfo{person}{Vikram Adve}.} \bibinfo{year}{2004}\natexlab{}.
\newblock \showarticletitle{LLVM: A compilation framework for lifelong program analysis \& transformation}. In \bibinfo{booktitle}{\emph{International symposium on code generation and optimization, 2004. CGO 2004.}} IEEE, \bibinfo{pages}{75--86}.
\newblock


\bibitem[Lautenbach et~al\mbox{.}(2021)]%
        {lautenbach2021proposing}
\bibfield{author}{\bibinfo{person}{Aljoscha Lautenbach}, \bibinfo{person}{Magnus Almgren}, {and} \bibinfo{person}{Tomas Olovsson}.} \bibinfo{year}{2021}\natexlab{}.
\newblock \showarticletitle{Proposing HEAVENS 2.0--an automotive risk assessment model}. In \bibinfo{booktitle}{\emph{Proceedings of the 5th ACM Computer Science in Cars Symposium}}. \bibinfo{pages}{1--12}.
\newblock


\bibitem[Litman(2017)]%
        {litman2017autonomous}
\bibfield{author}{\bibinfo{person}{Todd Litman}.} \bibinfo{year}{2017}\natexlab{}.
\newblock \showarticletitle{Autonomous vehicle implementation predictions}.
\newblock  (\bibinfo{year}{2017}).
\newblock


\bibitem[Liu et~al\mbox{.}(2021)]%
        {liu2021decision}
\bibfield{author}{\bibinfo{person}{Qi Liu}, \bibinfo{person}{Xueyuan Li}, \bibinfo{person}{Shihua Yuan}, {and} \bibinfo{person}{Zirui Li}.} \bibinfo{year}{2021}\natexlab{}.
\newblock \showarticletitle{Decision-making technology for autonomous vehicles: Learning-based methods, applications and future outlook}. In \bibinfo{booktitle}{\emph{2021 IEEE International Intelligent Transportation Systems Conference (ITSC)}}. IEEE, \bibinfo{pages}{30--37}.
\newblock


\bibitem[Liu et~al\mbox{.}(2022)]%
        {liu2022impact}
\bibfield{author}{\bibinfo{person}{Zongwei Liu}, \bibinfo{person}{Wang Zhang}, {and} \bibinfo{person}{Fuquan Zhao}.} \bibinfo{year}{2022}\natexlab{}.
\newblock \showarticletitle{Impact, challenges and prospect of software-defined vehicles}.
\newblock \bibinfo{journal}{\emph{Automotive Innovation}} \bibinfo{volume}{5}, \bibinfo{number}{2} (\bibinfo{year}{2022}), \bibinfo{pages}{180--194}.
\newblock


\bibitem[Livshits and Lam(2005)]%
        {livshits2005finding}
\bibfield{author}{\bibinfo{person}{V~Benjamin Livshits} {and} \bibinfo{person}{Monica~S Lam}.} \bibinfo{year}{2005}\natexlab{}.
\newblock \showarticletitle{Finding Security Vulnerabilities in Java Applications with Static Analysis.}. In \bibinfo{booktitle}{\emph{USENIX security symposium}}, Vol.~\bibinfo{volume}{14}. \bibinfo{pages}{18--18}.
\newblock


\bibitem[Lowe et~al\mbox{.}(2009)]%
        {lowe2009stride}
\bibfield{author}{\bibinfo{person}{Henry~J Lowe}, \bibinfo{person}{Todd~A Ferris}, \bibinfo{person}{Penni~M Hernandez}, {and} \bibinfo{person}{Susan~C Weber}.} \bibinfo{year}{2009}\natexlab{}.
\newblock \showarticletitle{STRIDE--An integrated standards-based translational research informatics platform}. In \bibinfo{booktitle}{\emph{AMIA annual symposium proceedings}}, Vol.~\bibinfo{volume}{2009}. \bibinfo{pages}{391}.
\newblock


\bibitem[Mao et~al\mbox{.}(2019)]%
        {mao2019conceptual}
\bibfield{author}{\bibinfo{person}{Xinyue Mao}, \bibinfo{person}{Mathias Ekstedt}, \bibinfo{person}{Engla Ling}, \bibinfo{person}{Erik Ringdahl}, {and} \bibinfo{person}{Robert Lagerstr{\"o}m}.} \bibinfo{year}{2019}\natexlab{}.
\newblock \showarticletitle{Conceptual abstraction of attack graphs-A use case of securiCAD}. In \bibinfo{booktitle}{\emph{International Workshop on Graphical Models for Security}}. Springer, \bibinfo{pages}{186--202}.
\newblock


\bibitem[Microsoft(2025)]%
        {microsoft_sbom_tool}
\bibfield{author}{\bibinfo{person}{Microsoft}.} \bibinfo{year}{2025}\natexlab{}.
\newblock \bibinfo{title}{SBOM Tool}.
\newblock
\urldef\tempurl%
\url{https://github.com/microsoft/sbom-tool}
\showURL{%
\tempurl}
\newblock
\shownote{Accessed: 2025-09-18}.


\bibitem[{MITRE}(2023a)]%
        {cwe119}
\bibfield{author}{\bibinfo{person}{{MITRE}}.} \bibinfo{year}{2023}\natexlab{a}.
\newblock \bibinfo{booktitle}{\emph{CWE-119: Improper Restriction of Operations within the Bounds of a Memory Buffer}}.
\newblock
\urldef\tempurl%
\url{https://cwe.mitre.org/data/definitions/119.html}
\showURL{%
\tempurl}


\bibitem[{MITRE}(2023b)]%
        {cwe120}
\bibfield{author}{\bibinfo{person}{{MITRE}}.} \bibinfo{year}{2023}\natexlab{b}.
\newblock \bibinfo{booktitle}{\emph{CWE-120: Buffer Copy without Checking Size of Input}}.
\newblock
\urldef\tempurl%
\url{https://cwe.mitre.org/data/definitions/120.html}
\showURL{%
\tempurl}


\bibitem[{MITRE}(2023c)]%
        {cwe126}
\bibfield{author}{\bibinfo{person}{{MITRE}}.} \bibinfo{year}{2023}\natexlab{c}.
\newblock \bibinfo{booktitle}{\emph{CWE-126: Buffer Over-read}}.
\newblock
\urldef\tempurl%
\url{https://cwe.mitre.org/data/definitions/126.html}
\showURL{%
\tempurl}


\bibitem[{MITRE}(2023d)]%
        {cwe134}
\bibfield{author}{\bibinfo{person}{{MITRE}}.} \bibinfo{year}{2023}\natexlab{d}.
\newblock \bibinfo{booktitle}{\emph{CWE-134: Use of Externally-Controlled Format String}}.
\newblock
\urldef\tempurl%
\url{https://cwe.mitre.org/data/definitions/134.html}
\showURL{%
\tempurl}


\bibitem[{MITRE}(2023e)]%
        {cwe190}
\bibfield{author}{\bibinfo{person}{{MITRE}}.} \bibinfo{year}{2023}\natexlab{e}.
\newblock \bibinfo{booktitle}{\emph{CWE-190: Integer Overflow or Wraparound}}.
\newblock
\urldef\tempurl%
\url{https://cwe.mitre.org/data/definitions/190.html}
\showURL{%
\tempurl}


\bibitem[{MITRE}(2023f)]%
        {cwe20}
\bibfield{author}{\bibinfo{person}{{MITRE}}.} \bibinfo{year}{2023}\natexlab{f}.
\newblock \bibinfo{booktitle}{\emph{CWE-20: Improper Input Validation}}.
\newblock
\urldef\tempurl%
\url{https://cwe.mitre.org/data/definitions/20.html}
\showURL{%
\tempurl}


\bibitem[{MITRE}(2023g)]%
        {cwe22}
\bibfield{author}{\bibinfo{person}{{MITRE}}.} \bibinfo{year}{2023}\natexlab{g}.
\newblock \bibinfo{booktitle}{\emph{CWE-22: Path Traversal}}.
\newblock
\urldef\tempurl%
\url{https://cwe.mitre.org/data/definitions/22.html}
\showURL{%
\tempurl}


\bibitem[{MITRE}(2023h)]%
        {cwe259}
\bibfield{author}{\bibinfo{person}{{MITRE}}.} \bibinfo{year}{2023}\natexlab{h}.
\newblock \bibinfo{booktitle}{\emph{CWE-259: Use of Hard-coded Password}}.
\newblock
\urldef\tempurl%
\url{https://cwe.mitre.org/data/definitions/259.html}
\showURL{%
\tempurl}


\bibitem[{MITRE}(2023i)]%
        {cwe327}
\bibfield{author}{\bibinfo{person}{{MITRE}}.} \bibinfo{year}{2023}\natexlab{i}.
\newblock \bibinfo{booktitle}{\emph{CWE-327: Use of a Broken or Risky Cryptographic Algorithm}}.
\newblock
\urldef\tempurl%
\url{https://cwe.mitre.org/data/definitions/327.html}
\showURL{%
\tempurl}


\bibitem[{MITRE}(2023j)]%
        {cwe330}
\bibfield{author}{\bibinfo{person}{{MITRE}}.} \bibinfo{year}{2023}\natexlab{j}.
\newblock \bibinfo{booktitle}{\emph{CWE-330: Use of Insufficiently Random Values}}.
\newblock
\urldef\tempurl%
\url{https://cwe.mitre.org/data/definitions/330.html}
\showURL{%
\tempurl}


\bibitem[{MITRE}(2023k)]%
        {cwe362}
\bibfield{author}{\bibinfo{person}{{MITRE}}.} \bibinfo{year}{2023}\natexlab{k}.
\newblock \bibinfo{booktitle}{\emph{CWE-362: Concurrent Execution using Shared Resource with Improper Synchronization ('Race Condition')}}.
\newblock
\urldef\tempurl%
\url{https://cwe.mitre.org/data/definitions/362.html}
\showURL{%
\tempurl}


\bibitem[{MITRE}(2023l)]%
        {cwe377}
\bibfield{author}{\bibinfo{person}{{MITRE}}.} \bibinfo{year}{2023}\natexlab{l}.
\newblock \bibinfo{booktitle}{\emph{CWE-377: Insecure Temporary File}}.
\newblock
\urldef\tempurl%
\url{https://cwe.mitre.org/data/definitions/377.html}
\showURL{%
\tempurl}


\bibitem[{MITRE}(2023m)]%
        {cwe400}
\bibfield{author}{\bibinfo{person}{{MITRE}}.} \bibinfo{year}{2023}\natexlab{m}.
\newblock \bibinfo{booktitle}{\emph{CWE-400: Uncontrolled Resource Consumption}}.
\newblock
\urldef\tempurl%
\url{https://cwe.mitre.org/data/definitions/400.html}
\showURL{%
\tempurl}


\bibitem[{MITRE}(2023n)]%
        {cwe502}
\bibfield{author}{\bibinfo{person}{{MITRE}}.} \bibinfo{year}{2023}\natexlab{n}.
\newblock \bibinfo{booktitle}{\emph{CWE-502: Deserialization of Untrusted Data}}.
\newblock
\urldef\tempurl%
\url{https://cwe.mitre.org/data/definitions/502.html}
\showURL{%
\tempurl}


\bibitem[{MITRE}(2023o)]%
        {cwe605}
\bibfield{author}{\bibinfo{person}{{MITRE}}.} \bibinfo{year}{2023}\natexlab{o}.
\newblock \bibinfo{booktitle}{\emph{CWE-605: Multiple Binds to the Same Port}}.
\newblock
\urldef\tempurl%
\url{https://cwe.mitre.org/data/definitions/605.html}
\showURL{%
\tempurl}


\bibitem[{MITRE}(2023p)]%
        {cwe703}
\bibfield{author}{\bibinfo{person}{{MITRE}}.} \bibinfo{year}{2023}\natexlab{p}.
\newblock \bibinfo{booktitle}{\emph{CWE-703: Improper Check or Handling of Exceptional Conditions}}.
\newblock
\urldef\tempurl%
\url{https://cwe.mitre.org/data/definitions/703.html}
\showURL{%
\tempurl}


\bibitem[{MITRE}(2023q)]%
        {cwe78}
\bibfield{author}{\bibinfo{person}{{MITRE}}.} \bibinfo{year}{2023}\natexlab{q}.
\newblock \bibinfo{booktitle}{\emph{CWE-78: OS Command Injection}}.
\newblock
\urldef\tempurl%
\url{https://cwe.mitre.org/data/definitions/78.html}
\showURL{%
\tempurl}


\bibitem[{MITRE}(2023r)]%
        {cwe807}
\bibfield{author}{\bibinfo{person}{{MITRE}}.} \bibinfo{year}{2023}\natexlab{r}.
\newblock \bibinfo{booktitle}{\emph{CWE-807: Reliance on Untrusted Inputs in a Security Decision}}.
\newblock
\urldef\tempurl%
\url{https://cwe.mitre.org/data/definitions/807.html}
\showURL{%
\tempurl}


\bibitem[{MITRE}(2023s)]%
        {cwe829}
\bibfield{author}{\bibinfo{person}{{MITRE}}.} \bibinfo{year}{2023}\natexlab{s}.
\newblock \bibinfo{booktitle}{\emph{CWE-829: Inclusion of Functionality from Untrusted Control Sphere}}.
\newblock
\urldef\tempurl%
\url{https://cwe.mitre.org/data/definitions/829.html}
\showURL{%
\tempurl}


\bibitem[{National Institute of Standards and Technology}(2022)]%
        {NIST2022}
\bibfield{author}{\bibinfo{person}{{National Institute of Standards and Technology}}.} \bibinfo{year}{2022}\natexlab{}.
\newblock \bibinfo{booktitle}{\emph{{Software Supply Chain Security Guidance Under Executive Order (EO) 14028 Section 4(e)}}}.
\newblock \bibinfo{type}{{T}echnical {R}eport}. \bibinfo{institution}{NIST}.
\newblock
\newblock
\shownote{Publication: February 4, 2022}.


\bibitem[Ohm et~al\mbox{.}(2020)]%
        {ohm2020towards}
\bibfield{author}{\bibinfo{person}{Marc Ohm}, \bibinfo{person}{Arnold Sykosch}, {and} \bibinfo{person}{Michael Meier}.} \bibinfo{year}{2020}\natexlab{}.
\newblock \showarticletitle{Towards detection of software supply chain attacks by forensic artifacts}. In \bibinfo{booktitle}{\emph{Proceedings of the 15th international conference on availability, reliability and security}}. \bibinfo{pages}{1--6}.
\newblock


\bibitem[Pendleton et~al\mbox{.}(2017)]%
        {pendleton2017perception}
\bibfield{author}{\bibinfo{person}{Scott~Drew Pendleton}, \bibinfo{person}{Hans Andersen}, \bibinfo{person}{Xinxin Du}, \bibinfo{person}{Xiaotong Shen}, \bibinfo{person}{Malika Meghjani}, \bibinfo{person}{You~Hong Eng}, \bibinfo{person}{Daniela Rus}, {and} \bibinfo{person}{Marcelo~H Ang}.} \bibinfo{year}{2017}\natexlab{}.
\newblock \showarticletitle{Perception, planning, control, and coordination for autonomous vehicles}.
\newblock \bibinfo{journal}{\emph{Machines}} \bibinfo{volume}{5}, \bibinfo{number}{1} (\bibinfo{year}{2017}), \bibinfo{pages}{6}.
\newblock


\bibitem[{PyCQA Bandit Developers}(2025)]%
        {banditTool}
\bibfield{author}{\bibinfo{person}{{PyCQA Bandit Developers}}.} \bibinfo{year}{2025}\natexlab{}.
\newblock \bibinfo{title}{Bandit: A security linter from PyCQA}.
\newblock \bibinfo{howpublished}{\url{https://bandit.readthedocs.io/en/latest/index.html}}.
\newblock
\newblock
\shownote{© Copyright 2025, Bandit Developers. Accessed: 2025-09-18}.


\bibitem[Raju et~al\mbox{.}(2019)]%
        {raju2019performance}
\bibfield{author}{\bibinfo{person}{Vysyaraju~Manikanta Raju}, \bibinfo{person}{Vrinda Gupta}, {and} \bibinfo{person}{Shailesh Lomate}.} \bibinfo{year}{2019}\natexlab{}.
\newblock \showarticletitle{Performance of open autonomous vehicle platforms: Autoware and Apollo}. In \bibinfo{booktitle}{\emph{2019 IEEE 5th International Conference for Convergence in Technology (I2CT)}}. IEEE, \bibinfo{pages}{1--5}.
\newblock


\bibitem[Ross(2012)]%
        {ross2012guide}
\bibfield{author}{\bibinfo{person}{Ronald~S Ross}.} \bibinfo{year}{2012}\natexlab{}.
\newblock \showarticletitle{Guide for conducting risk assessments}.
\newblock  (\bibinfo{year}{2012}).
\newblock


\bibitem[Saez-Perez et~al\mbox{.}(2025)]%
        {saez2025design}
\bibfield{author}{\bibinfo{person}{Javier Saez-Perez}, \bibinfo{person}{Julio Diez-Tomillo}, \bibinfo{person}{David Tena-Gago}, \bibinfo{person}{Jose~M Alcaraz-Calero}, {and} \bibinfo{person}{Qi Wang}.} \bibinfo{year}{2025}\natexlab{}.
\newblock \showarticletitle{Design, Implementation and Validation of a Level 2 Automated Driving Vehicle Reference Architecture}.
\newblock \bibinfo{journal}{\emph{Expert Systems}} \bibinfo{volume}{42}, \bibinfo{number}{6} (\bibinfo{year}{2025}), \bibinfo{pages}{e70050}.
\newblock


\bibitem[Semgrep(2025)]%
        {semgrep}
\bibfield{author}{\bibinfo{person}{Inc. Semgrep}.} \bibinfo{year}{2025}\natexlab{}.
\newblock \bibinfo{title}{Semgrep: Static Analysis for Code Security}.
\newblock
\urldef\tempurl%
\url{https://semgrep.dev/}
\showURL{%
\tempurl}
\newblock
\shownote{Accessed: 2025-09-18}.


\bibitem[Serebryany et~al\mbox{.}(2012)]%
        {serebryany2012addresssanitizer}
\bibfield{author}{\bibinfo{person}{Konstantin Serebryany}, \bibinfo{person}{Derek Bruening}, \bibinfo{person}{Alexander Potapenko}, {and} \bibinfo{person}{Dmitriy Vyukov}.} \bibinfo{year}{2012}\natexlab{}.
\newblock \showarticletitle{$\{$AddressSanitizer$\}$: A fast address sanity checker}. In \bibinfo{booktitle}{\emph{2012 USENIX annual technical conference (USENIX ATC 12)}}. \bibinfo{pages}{309--318}.
\newblock


\bibitem[Shepardson(2020)]%
        {Reuters2020}
\bibfield{author}{\bibinfo{person}{David Shepardson}.} \bibinfo{year}{2020}\natexlab{}.
\newblock \showarticletitle{Waymo opens driverless robo-taxi service to the public in Phoenix}.
\newblock \bibinfo{journal}{\emph{Reuters}} (\bibinfo{date}{Oct.} \bibinfo{year}{2020}).
\newblock
\urldef\tempurl%
\url{https://www.reuters.com/article/technology/waymo-opens-driverless-robo-taxi-service-to-the-public-in-phoenix-idUSKBN26T2Y3/}
\showURL{%
\tempurl}


\bibitem[Shladover(2018)]%
        {shladover2018connected}
\bibfield{author}{\bibinfo{person}{Steven~E Shladover}.} \bibinfo{year}{2018}\natexlab{}.
\newblock \showarticletitle{Connected and automated vehicle systems: Introduction and overview}.
\newblock \bibinfo{journal}{\emph{Journal of Intelligent Transportation Systems}} \bibinfo{volume}{22}, \bibinfo{number}{3} (\bibinfo{year}{2018}), \bibinfo{pages}{190--200}.
\newblock


\bibitem[Souppaya et~al\mbox{.}(2022)]%
        {souppaya2022secure}
\bibfield{author}{\bibinfo{person}{Murugiah Souppaya}, \bibinfo{person}{Karen Scarfone}, {and} \bibinfo{person}{Donna Dodson}.} \bibinfo{year}{2022}\natexlab{}.
\newblock \showarticletitle{Secure software development framework (ssdf) version 1.1}.
\newblock \bibinfo{journal}{\emph{NIST Special Publication}} \bibinfo{volume}{800}, \bibinfo{number}{218} (\bibinfo{year}{2022}), \bibinfo{pages}{800--218}.
\newblock


\bibitem[Team(2025)]%
        {ApolloCyberRT}
\bibfield{author}{\bibinfo{person}{Baidu~Apollo Team}.} \bibinfo{year}{2025}\natexlab{}.
\newblock \bibinfo{booktitle}{\emph{Apollo Cyber RT Framework}}.
\newblock
\urldef\tempurl%
\url{https://developer.apollo.auto/cyber.html}
\showURL{%
\tempurl}
\newblock
\shownote{Accessed on September 9, 2025}.


\bibitem[Teixeira et~al\mbox{.}(2024)]%
        {teixeira2024deterministic}
\bibfield{author}{\bibinfo{person}{Pedro~Veloso Teixeira}, \bibinfo{person}{Duarte Raposo}, \bibinfo{person}{Rui Lopes}, {and} \bibinfo{person}{Susana Sargento}.} \bibinfo{year}{2024}\natexlab{}.
\newblock \showarticletitle{Deterministic and Reliable Software-Defined Vehicles: key building blocks, challenges, and vision}.
\newblock \bibinfo{journal}{\emph{arXiv preprint arXiv:2407.17287}} (\bibinfo{year}{2024}).
\newblock


\bibitem[Teixeira et~al\mbox{.}(2020)]%
        {teixeira2020security}
\bibfield{author}{\bibinfo{person}{Rafael~R Teixeira}, \bibinfo{person}{Igor~P Maurell}, {and} \bibinfo{person}{Paulo~LJ Drews}.} \bibinfo{year}{2020}\natexlab{}.
\newblock \showarticletitle{Security on ROS: analyzing and exploiting vulnerabilities of ROS-based systems}. In \bibinfo{booktitle}{\emph{2020 Latin American robotics symposium (LARS), 2020 Brazilian symposium on robotics (SBR) and 2020 workshop on robotics in education (WRE)}}. IEEE, \bibinfo{pages}{1--6}.
\newblock


\bibitem[Ulbrich et~al\mbox{.}(2017)]%
        {ulbrich2017towards}
\bibfield{author}{\bibinfo{person}{Simon Ulbrich}, \bibinfo{person}{Andreas Reschka}, \bibinfo{person}{Jens Rieken}, \bibinfo{person}{Susanne Ernst}, \bibinfo{person}{Gerrit Bagschik}, \bibinfo{person}{Frank Dierkes}, \bibinfo{person}{Marcus Nolte}, {and} \bibinfo{person}{Markus Maurer}.} \bibinfo{year}{2017}\natexlab{}.
\newblock \showarticletitle{Towards a functional system architecture for automated vehicles}.
\newblock \bibinfo{journal}{\emph{arXiv preprint arXiv:1703.08557}} (\bibinfo{year}{2017}).
\newblock


\bibitem[Wahba(2024)]%
        {FORTUNE2024}
\bibfield{author}{\bibinfo{person}{Phil Wahba}.} \bibinfo{year}{2024}\natexlab{}.
\newblock \showarticletitle{Mercedes becomes the first automaker to sell autonomous cars in the U.S.}
\newblock \bibinfo{journal}{\emph{Fortune}} (\bibinfo{date}{April} \bibinfo{year}{2024}).
\newblock
\urldef\tempurl%
\url{https://fortune.com/2024/04/18/mercedes-self-driving-autonomous-cars-california-nevada-level-3-drive-pilot/}
\showURL{%
\tempurl}


\bibitem[Wang et~al\mbox{.}(2020)]%
        {wang2020bayesian}
\bibfield{author}{\bibinfo{person}{Jiali Wang}, \bibinfo{person}{Martin Neil}, {and} \bibinfo{person}{Norman Fenton}.} \bibinfo{year}{2020}\natexlab{}.
\newblock \showarticletitle{A Bayesian network approach for cybersecurity risk assessment implementing and extending the FAIR model}.
\newblock \bibinfo{journal}{\emph{Computers \& Security}}  \bibinfo{volume}{89} (\bibinfo{year}{2020}), \bibinfo{pages}{101659}.
\newblock


\bibitem[Wang et~al\mbox{.}(2019)]%
        {wang2019multi}
\bibfield{author}{\bibinfo{person}{Zhangjing Wang}, \bibinfo{person}{Yu Wu}, {and} \bibinfo{person}{Qingqing Niu}.} \bibinfo{year}{2019}\natexlab{}.
\newblock \showarticletitle{Multi-sensor fusion in automated driving: A survey}.
\newblock \bibinfo{journal}{\emph{Ieee Access}}  \bibinfo{volume}{8} (\bibinfo{year}{2019}), \bibinfo{pages}{2847--2868}.
\newblock


\bibitem[{Waymo LLC}(2024)]%
        {WaymoSite}
\bibfield{author}{\bibinfo{person}{{Waymo LLC}}.} \bibinfo{year}{2024}\natexlab{}.
\newblock \bibinfo{booktitle}{\emph{Waymo One: Autonomous rides in Phoenix, San Francisco, Los Angeles}}.
\newblock
\urldef\tempurl%
\url{https://waymo.com/rides/phoenix/}
\showURL{%
\tempurl}


\bibitem[Wheeler({[n.\,d.]})]%
        {wheelerFlawfinder}
\bibfield{author}{\bibinfo{person}{David~A. Wheeler}.} \bibinfo{year}{[n.\,d.]}\natexlab{}.
\newblock \bibinfo{title}{Flawfinder: A static analysis tool for C/C++}.
\newblock \bibinfo{howpublished}{\url{https://dwheeler.com/flawfinder/}}.
\newblock
\newblock
\shownote{Accessed: 2025-09-18}.


\bibitem[Wolf et~al\mbox{.}(2021)]%
        {wolf2021pasta}
\bibfield{author}{\bibinfo{person}{Andreas Wolf}, \bibinfo{person}{Dimitrios Simopoulos}, \bibinfo{person}{Luca D'Avino}, {and} \bibinfo{person}{Patrick Schwaiger}.} \bibinfo{year}{2021}\natexlab{}.
\newblock \showarticletitle{The PASTA threat model implementation in the IoT development life cycle}. In \bibinfo{booktitle}{\emph{INFORMATIK 2020}}. Gesellschaft f{\"u}r Informatik, Bonn, \bibinfo{pages}{1195--1204}.
\newblock


\bibitem[Xia et~al\mbox{.}(2023)]%
        {xia2023empirical}
\bibfield{author}{\bibinfo{person}{Boming Xia}, \bibinfo{person}{Tingting Bi}, \bibinfo{person}{Zhenchang Xing}, \bibinfo{person}{Qinghua Lu}, {and} \bibinfo{person}{Liming Zhu}.} \bibinfo{year}{2023}\natexlab{}.
\newblock \showarticletitle{An empirical study on software bill of materials: Where we stand and the road ahead}. In \bibinfo{booktitle}{\emph{2023 IEEE/ACM 45th International Conference on Software Engineering (ICSE)}}. IEEE, \bibinfo{pages}{2630--2642}.
\newblock


\bibitem[Xia et~al\mbox{.}(2025)]%
        {xia2025investigating}
\bibfield{author}{\bibinfo{person}{Lichen Xia}, \bibinfo{person}{Xing Gao}, {and} \bibinfo{person}{Weisong Shi}.} \bibinfo{year}{2025}\natexlab{}.
\newblock \showarticletitle{Investigating Security Threats in Multi-Tenant ROS 2 Systems}. In \bibinfo{booktitle}{\emph{2025 IEEE International Conference on Robotics and Automation (ICRA)}}. IEEE, \bibinfo{pages}{16441--16448}.
\newblock


\bibitem[Xiong et~al\mbox{.}(2019)]%
        {xiong2019threat}
\bibfield{author}{\bibinfo{person}{Wenjun Xiong}, \bibinfo{person}{Fredrik Krantz}, {and} \bibinfo{person}{Robert Lagerstr{\"o}m}.} \bibinfo{year}{2019}\natexlab{}.
\newblock \showarticletitle{Threat Modeling and Attack Simulations of Connected Vehicles: A Research Outlook.}. In \bibinfo{booktitle}{\emph{ICISSP}}. \bibinfo{pages}{479--486}.
\newblock


\bibitem[Ye et~al\mbox{.}(2023)]%
        {ye2023ros2}
\bibfield{author}{\bibinfo{person}{Yanlei Ye}, \bibinfo{person}{Zhenguo Nie}, \bibinfo{person}{Xinjun Liu}, \bibinfo{person}{Fugui Xie}, \bibinfo{person}{Zihao Li}, {and} \bibinfo{person}{Peng Li}.} \bibinfo{year}{2023}\natexlab{}.
\newblock \showarticletitle{Ros2 real-time performance optimization and evaluation}.
\newblock \bibinfo{journal}{\emph{Chinese Journal of Mechanical Engineering}} \bibinfo{volume}{36}, \bibinfo{number}{1} (\bibinfo{year}{2023}), \bibinfo{pages}{144}.
\newblock


\bibitem[Yurtsever et~al\mbox{.}(2020)]%
        {yurtsever2020survey}
\bibfield{author}{\bibinfo{person}{Ekim Yurtsever}, \bibinfo{person}{Jacob Lambert}, \bibinfo{person}{Alexander Carballo}, {and} \bibinfo{person}{Kazuya Takeda}.} \bibinfo{year}{2020}\natexlab{}.
\newblock \showarticletitle{A survey of autonomous driving: Common practices and emerging technologies}.
\newblock \bibinfo{journal}{\emph{IEEE access}}  \bibinfo{volume}{8} (\bibinfo{year}{2020}), \bibinfo{pages}{58443--58469}.
\newblock


\bibitem[Zalewski(2017)]%
        {zalewski2017american}
\bibfield{author}{\bibinfo{person}{Michal Zalewski}.} \bibinfo{year}{2017}\natexlab{}.
\newblock \bibinfo{title}{American fuzzy lop}.
\newblock


\bibitem[Zhao et~al\mbox{.}(2024)]%
        {zhao2024fuzzer}
\bibfield{author}{\bibinfo{person}{Xiaoqi Zhao}, \bibinfo{person}{Haipeng Qu}, \bibinfo{person}{Jiaohong Yi}, \bibinfo{person}{Jinlong Wang}, \bibinfo{person}{Miaoqing Tian}, {and} \bibinfo{person}{Feng Zhao}.} \bibinfo{year}{2024}\natexlab{}.
\newblock \showarticletitle{A fuzzer for detecting use-after-free vulnerabilities}.
\newblock \bibinfo{journal}{\emph{Mathematics}} \bibinfo{volume}{12}, \bibinfo{number}{21} (\bibinfo{year}{2024}), \bibinfo{pages}{3431}.
\newblock


\end{thebibliography}


\end{document}